# Designing all graphdiyne materials as graphene derivatives: topologically driven modulation of electronic properties

Patrick Serafini[1], Alberto Milani[1*], Davide M. Proserpio[2,3], Carlo S. Casari[1*]

[1]*Dipartimento di Energia, Politecnico di Milano, via Ponzio 34/3, Milano, Italy*
[2]*Dipartimento di Chimica, Università degli Studi di Milano, 20133 Milano, Italy*
[3]*Samara Center for Theoretical Materials Science (SCTMS), Samara State Technical University, Samara 443100, Russia*

*Corresponding authors: alberto.milani@polimi.it; carlo.casari@polimi.it

**Abstract**

Designing new 2D systems with tunable properties is an important subject for science and technology. Starting from graphene, we developed an algorithm to systematically generate 2D carbon crystals belonging to the family of graphdiynes (GDYs) and having different structures and sp/sp2 carbon ratio. We analyze how structural and topological effects can tune the relative stability and the electronic behavior, to propose a rationale for the development of new systems with tailored properties. A total of 26 structures have been generated, including the already known polymorphs such as α-, β- and γ-GDY. Periodic density functional theory calculations have been employed to optimize the 2D crystal structures and to compute the total energy, the band structure, and the density of states. Relative energies with respect to graphene have been found to increase when the values of carbon sp/sp2 ratio increase, following however different trends based on the peculiar topologies present in the crystals. These topologies also influence the band structure giving rise to semiconductors with a finite bandgap, zero-gap semiconductors displaying Dirac cones, or metallic systems. The different trends allow identifying some topological effects as possible guidelines in the design of new 2D carbon materials beyond graphene.

## 1. Introduction

Carbon materials and their nanostructures played a relevant role in the science and technology of the last two decades: from fullerenes to carbon nanotubes, from polyconjugated polymers to graphene, the so-called "era of carbon allotropes" has been enlightened by groundbreaking results and Nobel prizes, paving the way to many interesting research topics [1]. In the last years, the interest of many scientists has been directed towards exotic forms of carbon, including systems based on 1D sp-hybridized carbon (variously referred to as carbyne, carbon-atom wires, polyynes, cumulenes, …) and on hybrid sp-sp$^2$ carbon systems. These investigations focused both on the fundamental properties and the potential applications in different fields, showing promising perspectives for the near future [2-6]. Graphyne (GY) and graphdiyne (GDY) represent 2D carbon crystals with sp-sp2 carbon atoms [7-14]. They can be constructed as possible modification of graphene by interconnecting sp$^2$-carbon hexagons with linear sp-carbon chains of different lengths (a single or a double acetylenic bond for GY and GDY, respectively), generating new systems with peculiar and tunable electronic and optical properties. Moreover, starting from GY and GDY, many other ideal 2D hybrid sp-sp$^2$ carbon systems can be proposed by playing with geometry and topology offering countless possibilities in the design and tailoring of carbon allotropes, both theoretically and experimentally.

Early theorethical studies on GY and GDY-based system were reported in 1987 [15] and recently with modern computational methods to shed light on their properties [16-18]. These structures are a part of a larger family of two-dimensional π-conjugated Covalent Organic Frameworks (COF) showing the occurrence of Dirac cones [19, 20], flat bands, tunable bands gap. Such phenomena observed in the electronic structure of COFs have been explained based on peculiar topological effects also in connection to their influence on the charge transport behavior [21-24].

Several papers report on the prediction of properties of γ-GDY mainly through Density Functional Theory (DFT) calculations [5,6,25-28,29]. From the experimental side, synthetic bottom-up approaches have been successfully employed to produce sub-fragments of GDY of different topology and dimensions in particular by Haley and coworkers [30-34]. Later, different papers reported the preparation and characterization of extended 2D GDY sheets prepared through organometallic synthesis techniques, showing promising routes to the production of these systems, even though significant efforts should still be put to further investigate and understand their properties [35-37]. Recent advances in on-surface synthesis allow the preparation of hybrid sp-sp$^2$ carbon nanostructures and their atomic-scale investigation by surface science techniques [8][36][38-40].

The recent possibilities to realize new systems have opened new opportunities well beyond the sole investigation of their fundamental properties. Hence, the identification of possible guidelines to support the synthetic efforts is mandatory in a knowledge-based research approach. In addition, to develop a new class of 2D carbon materials some relevant open issues still need to be addressed. First, how many different 2D crystal structures can be possible? What about their stability? Which structures are metallic, semimetallic with Dirac cones or semiconducting? Is there any relation between the crystal structure and the electronic behaviour?

To answer these questions, we have developed an algorithm to systematically generate new hybrid sp-sp$^2$ carbon structures as modifications of graphene by introducing linear diacetylenic units. By DFT calculations on the geometries so generated, we performed geometry optimization, evaluation of the relative stability, and prediction of electronic band structure, gap and Density of States (DOS). We analysed a total of 26 structures more than half not previously identified and we outlined metallic, semimetallic with Dirac cones and semiconducting systems grouped on the basis of topological features. The identification of new 2D carbon structures and the topology based electronic properties give further insight for the design and understanding of new hybrid sp-sp$^2$ carbon 2D materials.

## 2. Theoretical Details

To identify systematically all the sp-sp$^2$ carbon systems in the graphdyine family, we used ToposPro [41] to generate subnets of graphene where bonds are deleted in all possible ways. This procedure was used in the past to generate uninodal and binodal nets [42,43]. Based on the 2D crystal structures selected by ToposPro, Periodic Boundary Conditions (PBC) DFT simulations have been carried out by employing CRYSTAL17 [44, 45] to optimize the geometry (both atomic position and cell parameters) and compute the electronic band structure and DOS. To this aim, we adopted the PBE0 hybrid exchange-correlation functionals together with 6-31G(d) gaussian basis sets [46]. This level of theory has been chosen according to our previous investigations of the structural and vibrational properties of γ-GDY and related nanoribbons, where the results obtained by using different functionals and basis sets have been compared [29]. When using the 6-31G(d) basis set in PBC-DFT simulations with the CRYSTAL code, the exponent of the diffuse sp orbitals of carbon atoms have been increased from 0.1687144 Bohr$^{-2}$ to 0.187 Bohr$^{-2}$ to avoid convergence problems in the SCF, due to basis sets linear dependencies [47]. Considering the other simulation parameters, the tolerance on integral screening has been fixed to 9,9,9,9,80 (TOLINTEG parameters), while the shrink parameters defining Monkhorst-Pack and Gilat sampling points have been fixed to 100 and 200 for the calculation of the band structure and DOS, respectively. Depending on the crystalline structure (orthorhombic, monoclinic and hexagonal), the main three paths and special points in the Brillouin zone were chosen. Band structures and DOS were plotted using the program CRYSPLOT, a visualization environment for plotting properties of crystalline solids as computed through the CRYSTAL code (http://crysplot.crystalsolutions.eu/).

Similar to ref. [21], the data here reported have been obtained by using PBE0 functional, taking advantages of the improvement obtained by means of hybrid functional in the description of ground state electronic properties. Even if a further improvement could be obtained by employing the HSE06 functional [48], in our previous investigation on γ-GDY and its nanoribbons we verified that both PBE0 and HSE06 are able to describe the same trends for band gaps, with a larger overestimation of PBE0's ones with respect to benchmark values computed by the GW method [29]. For some peculiar structures, full geometry optimization, band structure and DOS calculations have been carried out also by HSE06 functional and compared with PBE0 results.

## 3. Results and Discussion

### 3.1 Construction of graphdiyne crystals as graphene derivatives

We developed an approach to generate and classify all possible graphdiyne 2D structures. We considered stable graphene derivatives by inserting linear diacetylene ($C_4$) groups for a given number of 3-coordinated carbon atoms (sp$^2$ like) per primitive cell. We limited the analysis to maximum 8 sp$^2$ carbon atoms per primitive cell to avoid too large cells, and to focus on structures that are more likely to be experimentally synthesized. This limit is enough to find all previously reported and many other GDY-like structures. Our approach is based on removing edges from the graphene structure and substituting them with linear diacetylenic units. The starting set of honeycomb layers with deleted edges was made of about 40000 structures with a maximum of 8 carbon atoms per primitive cell (4 times the original cell). With the help of the topological classification tools in ToposPro we extracted 332 topologically distinct patterns containing 6-membered rings with one or more missing edges. Geometrical considerations led to the possible derived hexagons with deleted edges (from one to six) as shown in Figure 1. For each configuration there is a "dual" one with reverse deleted edges, leaving 12 possible distinct patterns. In Figure 1-left we show the eight patterns that after the introduction $C_4$ edges will not allow the closure of the 6-membered ring without a large angular distortion (see the detail for one case). Only the four remaining configurations shown in Figure 1-right allow expansion by the insertion of $C_4$ linear structures without substantial distortion, i.e. keeping all the angles around 120°. These four configurations and the original unaltered aromatic ring constitute five building blocks, here called h,H,R,T,L (small and large hexagon h and H, rhombus R, triangle T and line L), as suggested by Park et al. [17]. From the 332 patterns we extracted only those containing some of the five possible building blocks that allow to tile the plane without large distortion, obtaining 26 structures GDY-like (of which 17 are new) described in Figure 2 and Table 1. Due to the "dual" properties of L/R and h/H the 26 structures can be grouped in 12 couples plus a self-dual layer. Based on this classification structures are called $6\text{-}h^nL^mT^oR^pH^q$, where 6 represents the number of carbon atoms along the longest edge and the superscript on the building block symbols represents the number of each block appearing in a primitive unit cell (for example the primitive cell of *c*2*mm* 6-h$^2$L contains two small $C_6$ h-rings and one $C_{14}$ L-ring and all the edges are either between 2 or 6 carbon atoms).

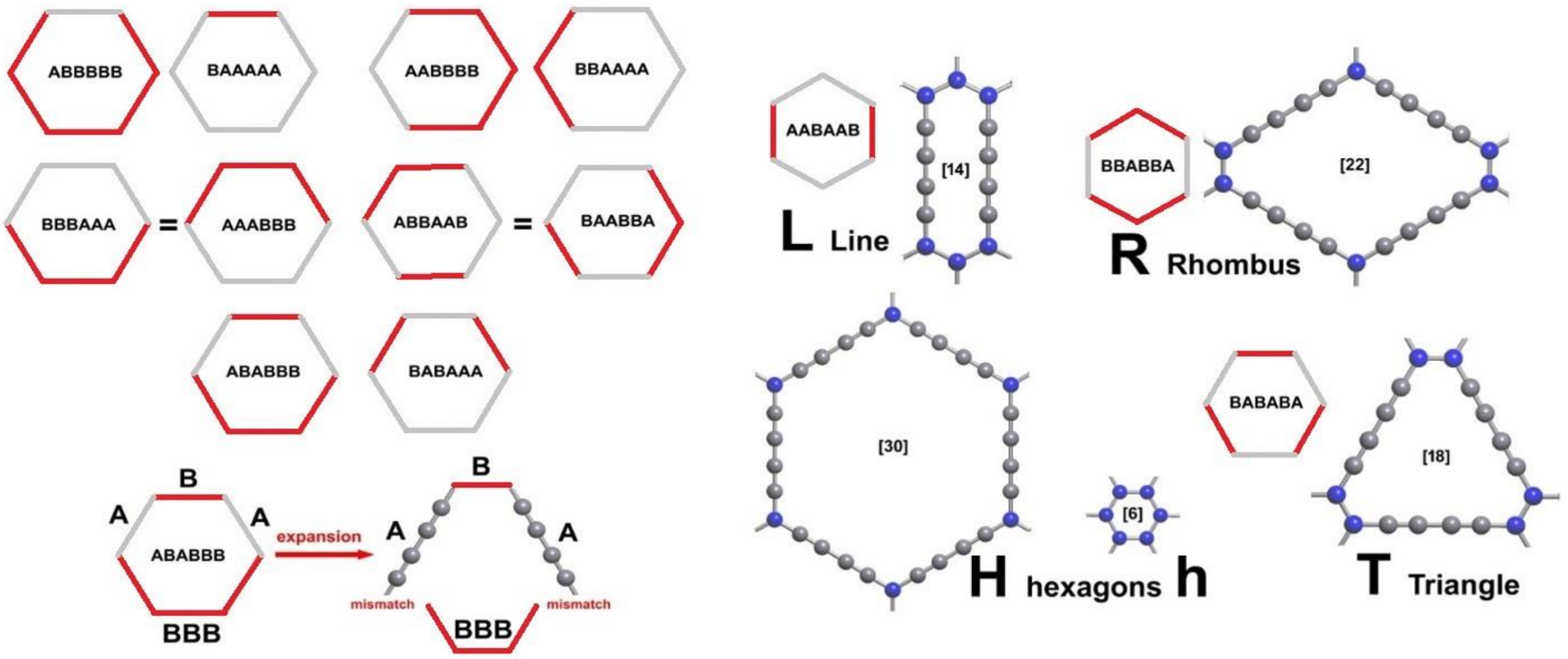

***Figure 1***: Left) The eight distinct possible configurations that upon expansion give highly distorted hexagons that are discarded in our generation of GDY-like layers. A and B indicates non-equivalent sides of the hexagon units: B should be considered as a true CC bond, while A represent the diacetylene unit ($C_4$) of four sp carbon atoms. The configurations are shown as "dual" couples, i.e. reversing the role of A vs. B. Two couples do not generate a new configuration, e.g. BBBAAA=AAABBB. Right) The four configurations that upon expansion gave undistorted six-sided polygons. The value in square bracket is the total number of carbon atom in the ring that uniquely define each building block h=[6], H=[30]; L=[14]; T=[18], R=[22]. Line (L) and rhombus (R) blocks are "dual", so for each layer containing them we can substitute each L with an R without significant distortions. The definition here proposed follows the one reported in the work by Park et al. [17]

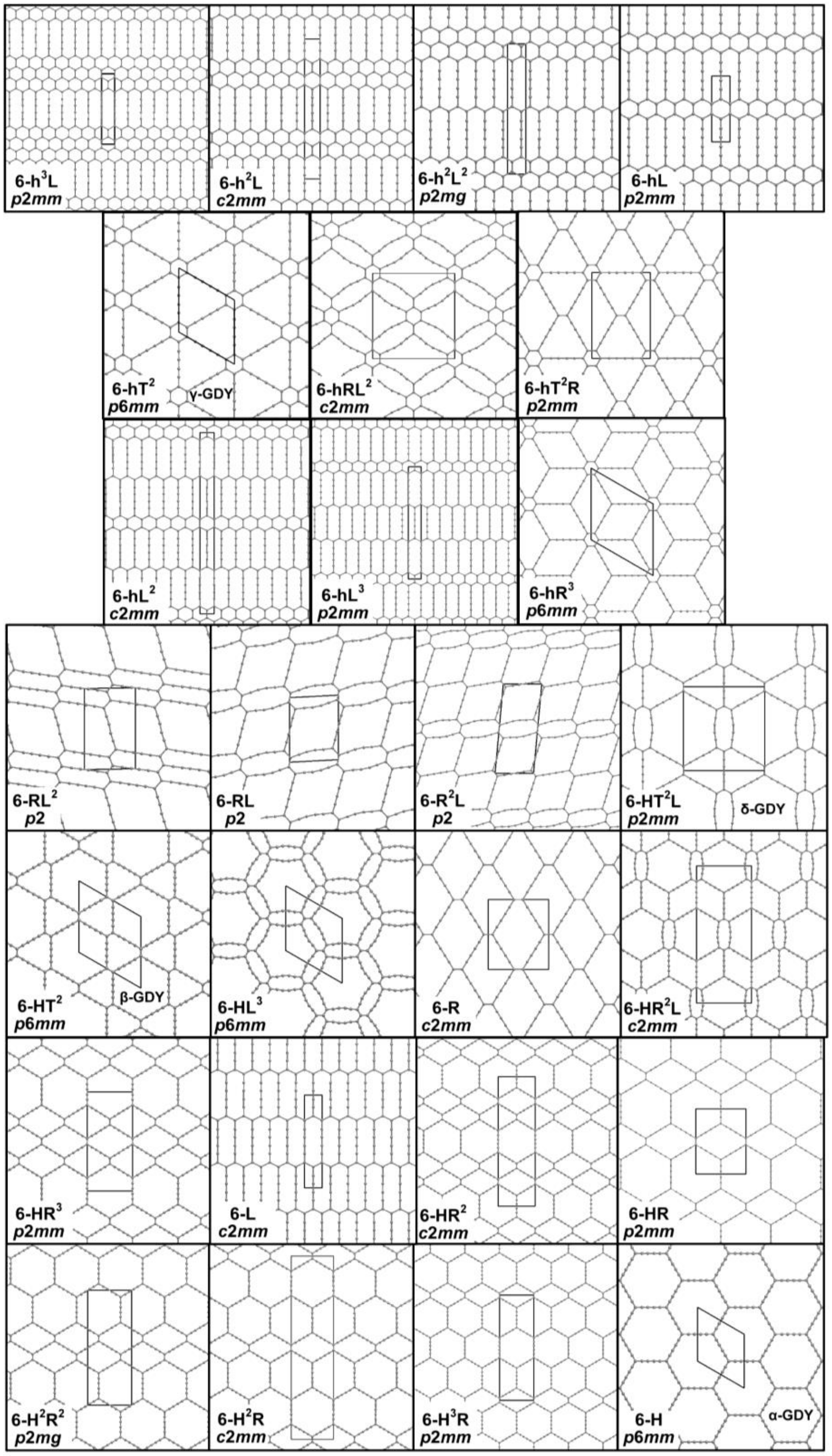

***Figure 2:*** The 26 2D-structures identified and investigated in this work.
The unit cells and the plane groups are indicated. α-,β,γ-,δ-GDY are also labelled

***Table 1***: Summary of the 26 structures here investigated reporting name, relative energy with respect to graphene, plane group, Pearson symbol, $sp/sp^2$ ratio and electronic character (band gap reported for semiconductors). In the last column, the reference number of previous papers investigating the same structure are reported [16-18].

| Name | Rel. Energy kcal/mol | Plane group | Pearson Symbol | $sp/sp^2$ ratio | Electronic Character (PBE0) | Refs. |
|---|---|---|---|---|---|---|
| 6-$h^3$L | 14.15 | *p2mm* | oP12 | 0.50 | metal | This work |
| 6-$h^2$L | 16.81 | *c2mm* | oS20 | 0.66 | metal | This work |
| 6-$h^2L^2$ | 20.55 | *p2mg* | oP16 | 1.00 | metal | This work |
| 6-hL | 20.59 | *p2mm* | oP8 | 1.00 | metal | This work |
| 6-$hT^2$ γ-GDY | 21.13 | *p6mm* | hP18 | 2.00 | B.G.=1.63 | [16],[17],[18] |
| 6-$hRL^2$ | 22.65 | *c2mm* | oS48 | 2.00 | 0 B.G. | This work |
| 6-$hT^2$R | 22.69 | *p2mm* | oP28 | 2.50 | B.G.=0.83 | [17],[18] |
| 6-$hL^2$ | 23.11 | *c2mm* | oS28 | 1.33 | metal | This work |
| 6-$hL^3$ | 24.06 | *p2mm* | oP20 | 1.50 | metal | This work |
| 6-$hR^3$ | 24.07 | *p6mm* | hP32 | 3.00 | 0 B.G. | [17],[18] |
| 6-$RL^2$ | 24.73 | *p2* | mP22 | 2.66 | 0 B.G. | This work |
| 6-RL | 24.73 | *p2* | mP16 | 3.00 | 0 B.G. | This work |
| 6-$R^2$L | 25.03 | *p2* | mP26 | 3.33 | 0 B.G. | This work |
| 6-$HT^2$L δ-GDY | 25.07 | *p2mm* | oP36 | 3.50 | B.G.=0.18 | This work |
| 6-$HT^2$ β-GDY | 25.40 | *p6mm* | hP30 | 4.00 | B.G.=1.14 | [17],[18] |
| 6-$HL^3$ | 25.41 | *p6mm* | hP32 | 3.00 | 0 B.G. | [18] |
| 6-R | 25.48 | *c2mm* | oS20 | 4.00 | 0 B.G. | [16],[17],[18] |
| 6-$HR^2$L | 25.85 | *c2mm* | oS80 | 4.00 | 0 B.G. | This work |
| 6-$HR^3$ | 26.11 | *p2mm* | oP44 | 4.50 | 0 B.G. | This work |
| 6-L | 26.18 | *c2mm* | oS12 | 2.00 | metal | [16],[18] |
| 6-$HR^2$ | 26.30 | *c2mm* | oS68 | 4.66 | 0 B.G. | This work |
| 6-HR | 26.63 | *p2mm* | oP24 | 5.00 | 0 B.G. | [17],[18] |
| 6-$H^2R^2$ | 26.63 | *p2mg* | oP48 | 5.00 | 0 B.G. | This work |
| 6-$H^2$R | 26.92 | *c2mm* | oS76 | 5.33 | 0 B.G. | This work |
| 6-$H^3$R | 27.05 | *p2mm* | oP52 | 5.50 | 0 B.G. | This work |
| 6-H | 27.39 | *p6mm* | hP14 | 6.00 | 0 B.G. | [16],[17],[18] |

### 3.1 Relative energies of 2D crystals

After the full geometry optimization of the 26 structures and of the reference 2D graphene structure, we investigated their stability by calculating the relative energy per carbon atom with respect to graphene:

$$E_{rel}^{i} = \frac{E_{tot}^{i}}{N^{i}} - \frac{E_{tot}^{graphene}}{N^{graphene}}$$

where $E_{tot}$ is the DFT-computed total energy and $N$ is the number of atoms in the unit cell for the $i$-th structure and graphene respectively ($N^{graphene} = 2$). This value of $R_{rel}$ gives the relative cohesive energy per carbon atom and allows to identify the most stable structures. Relative energy values are plotted as a function of the sp/sp$^2$ ratio, calculated as the ratio of sp and sp$^2$ carbon atoms numbers in the unit cell (see Figure 3). This ratio ranges from 0 in the case of graphene up to 6 in the case of $\alpha$-GDY, the largest values possible for the periodic 2D hybrid sp/sp$^2$ carbon nanostructures here investigated where the sp domains are formed by diacetylenic units. The numerical values of the total and relative energies, sp/sp$^2$ ratios and layer densities are reported for all the structures in the Supporting Information (Table S2).

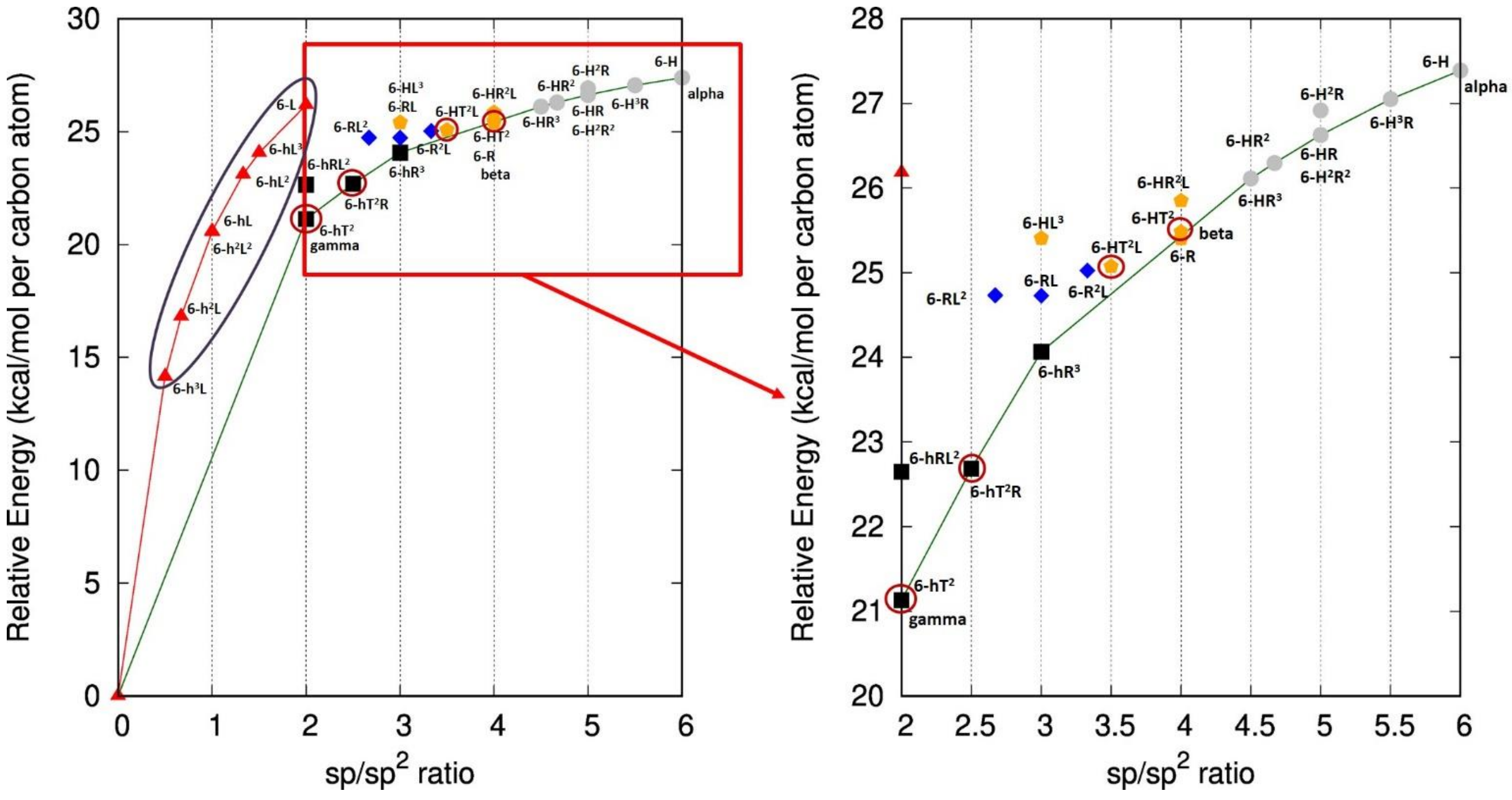

***Figure 3:*** Plots of the DFT computed relative energies of the 26 structures here investigated with respect to graphene as a function of sp-sp$^2$ ratios. In the first panel the two distinct trends are analyzed while in the second panel, the members of the second trends are collected in different classes, depending on the peculiar topology identified. The seven metallic systems are all in the left panel and grouped in the elongated ellipsoid while semiconductors with finite band gaps are circled in red.

The increasing amount of sp carbon atoms with respect to graphene (i.e. a larger sp/sp$^2$ ratio) increases the energy of the whole system, consistently with chemical intuition. However, in Figure 3 two different trends can be identified. The first trend (red line) is continuous and smooth and contains all the structures formed by different combinations of 6-atom hexagons (h) and line polygons (L) and namely 6-h$^3$L, 6-h$^2$L, 6-hL, 6-h$^2$L$^2$, 6-hL$^2$, 6-hL$^3$ and 6-L and graphene (6-h). The lowest energy structure is 6-h$^3$L, formed by graphene ribbons having a width of 3 aromatic units and interconnected by diacetylenic bridges. 6-h$^2$L and 6-hL are formed by graphene ribbons having a width of 2 and 1 aromatic units respectively and interconnected by diacetylenic bridges show increase of relative energy. Hence, by reducing the width of graphene ribbons (i.e. as far as we progressively move away from the graphene limit) an increase of the relative energy is observed. Similarly, a further increase in energy is found in 6-

$hL^2$ and 6-$hL^3$ which have, as in 6-hL, graphene nanoribbons of width 1 but which are interconnected by 2 and 3 L polygons (bridges of two and three diacetylenic units), respectively. Therefore, both the number of condensed aromatic hexagons (width of graphene ribbon) and the spacing between the graphene ribbons (h polygons) set by the number of diacetylenic domains (L polygons) modulates the relative energy of these systems with respect to graphene. The maximum energy is reached in 6-L where no h units are present at all. Considering therefore the general trend given by structures 6$h^nL^m$ in which graphene nanoribbons of width "n" are connected by "m" diacetylenic units, the lower energy is found by increasing n and decreasing m, clearly tending closer and closer to the graphene case. Interestingly, these two parameters seem to have similar weight, in fact the same energy is obtained in 6-hL with the smallest graphene ribbon and single diacetylenic unit and in 6-$h^2L^2$ in which the increase of energy given by doubling the diacetylenic units is counterbalanced by doubling the graphene ribbon width. The structures of these group have been investigated in a recent work with the name "grazyines " [49]. Our analysis underlines a clear trend in the energy that can be easily generalized for similar structures for predictive purposes.

A different trend in Figure 3 is indicated by the green curve. In this group we find the widely studied polymorphs of GDY, usually labeled as α-GDY (structure 6-H), β-GDY (structure 6-$HT^2$) and γ-GDY (structure 6-$hT^2$). γ-GDY has gathered more attention in the recent years, also from the experimental point of view. Among all the structures belonging to this trend, it has the lowest energy with respect to graphene, consistently with its low sp/$sp^2$ ratio (sp/sp2=2). On the other hand, α-GDY is the highest energy structure among all of those which are investigated here, again consistently with the highest sp/$sp^2$ ratio (sp/sp2=6). β-GDY with sp/sp2=4 is in between these two limiting cases.

Apart from these widely studied polymorphs, the structures belonging to this second group can be further classified based on their topology and structure. The description of their geometry is based on their building units, similarly to the method proposed by Park et al.[17]. As shown in Figure 3, the four lowest energy structures 6-$hT^2$ (γ-GDY), 6-$hRL^2$, 6-$hT^2R$, 6-$hR^3$ form one subgroup themselves since they all share the presence of h polygons in their geometry, i.e. a last reminiscence of the graphene structure. On the other hand, the highest energy structures 6-$HR^3$, 6-$HR^2$, 6-HR, 6-$H^2R^2$, 6-$H^2R$, 6-$H^3$, 6-H(α-GDY) form another subgroup and they can be all described as α-GDY ribbons connected with different widths and interconnections, tending indeed towards the upper limit of α-GDY.

In the intermediate case, two clusters of structures can be identified: one collects structures 6-$HL^3$, 6-$HT^2L$, 6-R, 6-$HR^2L$ and 6-$HT^2$ (β-GDY) and the other one the structures 6-RL, 6-$RL^2$ and 6-$R^2L$. The members in the first group are all characterized by the presence of isolated α-GDY units (i.e. large 30-atoms hexagons H with all sizes characterized by diacetylenic units) while in the second group the systems are structurally peculiar, since they are formed only by units which can be described as rombus R and lines L according to definition in Figure 1. It should be noticed that in many of the structures containing L units, the diacetylenic bridges are usually bent and not linear, as also found in the calculations by Belenkov et al. [16].

The energy trend can be summarized in the following: 1) As expected, the relative energy with respect to graphene increases for increasing sp/$sp^2$ ratio. 2) The closer is the structural resemblance to graphene, the lower is the relative energy of the 2D structures. This is clearly shown in crystals made by graphene ribbons having diacetylenic connected bridges: the more we approach the graphene structures (the lower is the sp/$sp^2$ ratio), the lower is the relative energy. 3) Considering the other trends, the lowest energy structures are those where the graphene h unit (hexagon of $sp^2$ carbon) is still present in the structure. 4) The energy increases when the number of α-GDY units (H hexagon) increases up to the limiting case of 2D α-GDY.

These trends in relative energy are consistent with the cohesive energies reported by Park et al.

[17] and with the sublimation energy given by Belenkov et al. [16] by considering that a larger relative energy corresponds to a lower cohesive and sublimation energy. Our work consistently with Park et al. [17] shows that α-GDY has the lowest stability while γ-GDY is the most stable. On the other hand, the semiempirical calculations by Belenkov et al. [16] predict the largest sublimation energy (i.e. lower relative energy) for the system called γ2- , which corresponds to 6-L in our work. For 6-L, we find a larger relative energy than γ-GDY even if they share the same value of $sp/sp^2$ ratio. Our results shows a general qualitative agreement with Ref. [18]. Relative energy (here with respect to graphene, in [18] with respect to γ-graphyne) increases with decreasing carbon densities (i.e. increasing $sp/sp^2$ ratio). Our work includes some structures already investigated in the literature and several novel ones, thus revealing the wide range of possibilities available in the design of 2D $sp$-$sp^2$ carbon systems.

In Fig. 3 seven structures that show a metallic behaviour are grouped with an elongated ellipsoid while the finite band gap semiconducting crystals are circled in red. All the other 2D carbon systems are 0 band gap semiconductors, as discussed in the following chapter.

### 3.2 Electronic properties: band structure and DOS

We here discuss how topological elements affect the band structure and the electronic properties in graphdiyne crystals, on the basis of their peculiar crystal structure. In Figure 4 we compare the band structure of graphene with the three widely investigated α-, β- and γ- polymorphs of GDY. α-GDY shows very similar band structure to graphene, with the occurrence of a Dirac Cones at K-point, which makes both of them zero-gap semiconductors or semi-metals, in agreement with previous works [19,20]. Graphene and α-GDY have the same crystal structure and layer group (p6mm) differing only in the number of carbon-atoms in their basis (2 and 14 in in graphene and α-GDY, respectively), highlighting the topology-dependence of band structure. In α-GDY we also observed the appearance of both occupied and empty flat bands, another typical topology-dependent feature found for 2D COFs [21-24]. On the other hand, β- and γ-GDY are finite-gap semiconductors, showing a band gap of 1.14 and 1.63 eV, respectively. Their structures are similar: based on the algorithm we adopted to build the structures, β- and γ-GDY are indeed dual one to the other (see also Figures 1 and 2), respectively 6-HT$^2$ and 6hT$^2$, demonstrating a topology-dependence of the band structure.

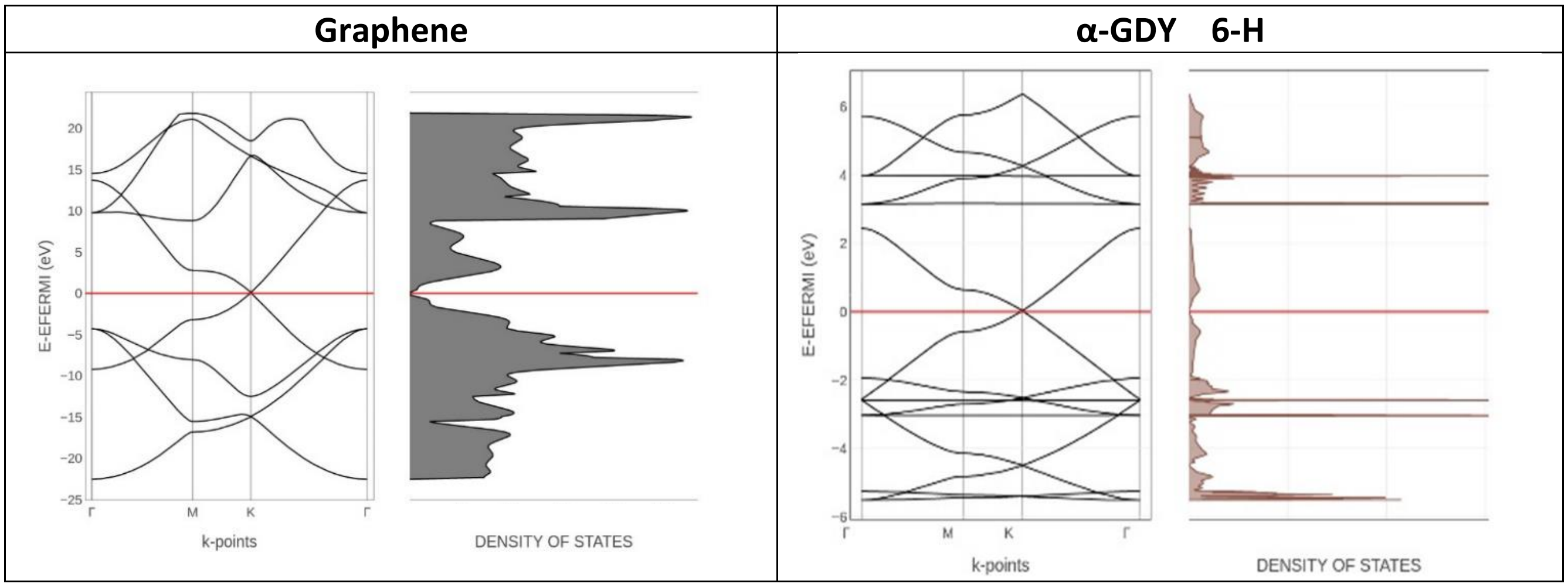

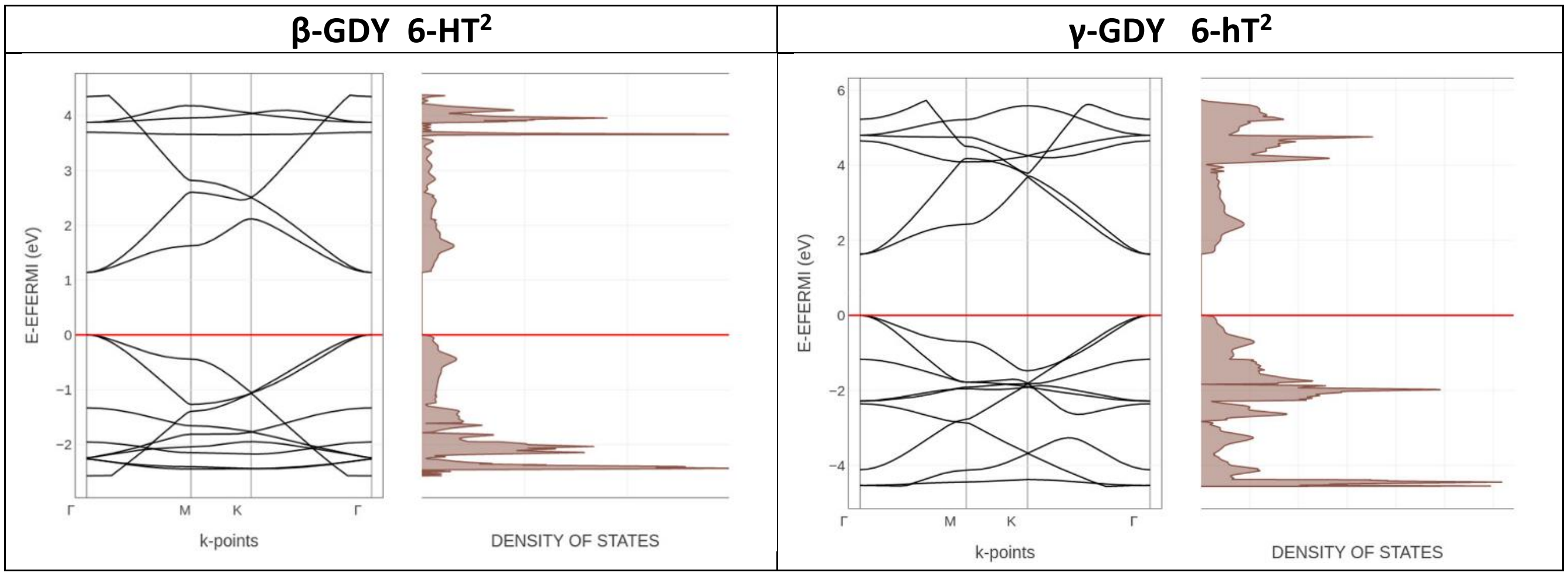

***Figure 4:*** Comparison of DFT computed band structures and DOS of graphene and α-, β- and γ-GDY polymorphs.

Further results can be obtained by extending the analysis to all the systems here investigated. The band structure and DOS of all these structures can be classified into three classes: finite-gap semiconductors (such as β- and γ-GDY), metals and zero-gap semiconductors (see Figure 5). For each class, the band structure and DOS of only one representative structure is reported, while a complete table for all the geometries investigated is reported in the Supporting Information (Table S1).

Finite-gap semiconductors include, β- and γ-GDY, and only two other structures, 6-hT$^2$R and 6-HT$^2$L, presenting a finite gap of 0.8327 and 0.1789 eV, respectively. These geometries are dual one with respect to the other, as for β- and γ-GDY (6-HT$^2$ and 6-hT$^2$ respectively, see Figure 1). Among the 26 structures here investigated, these ones are the only containing a T shaped unit (see Figure 1), suggesting that T units would induce electronic effects leading to the occurrence of a band gap.

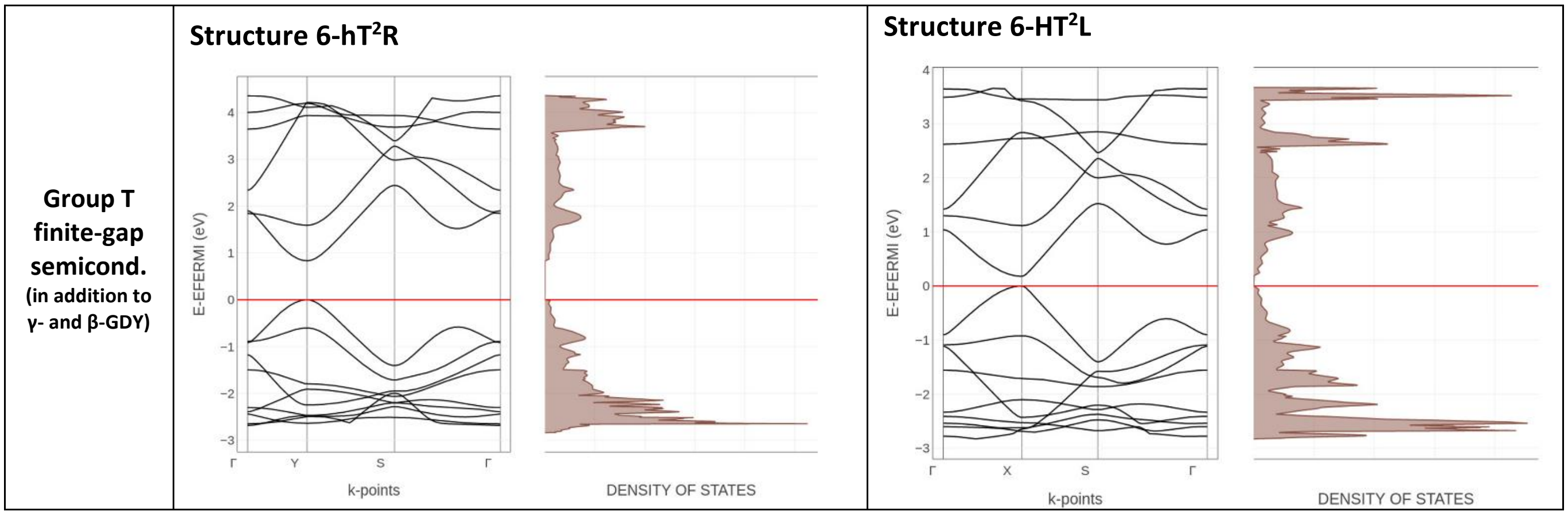

| | | |
|---|---|---|
| **metals** | **Structure 6-h$^2$L$^2$**<br>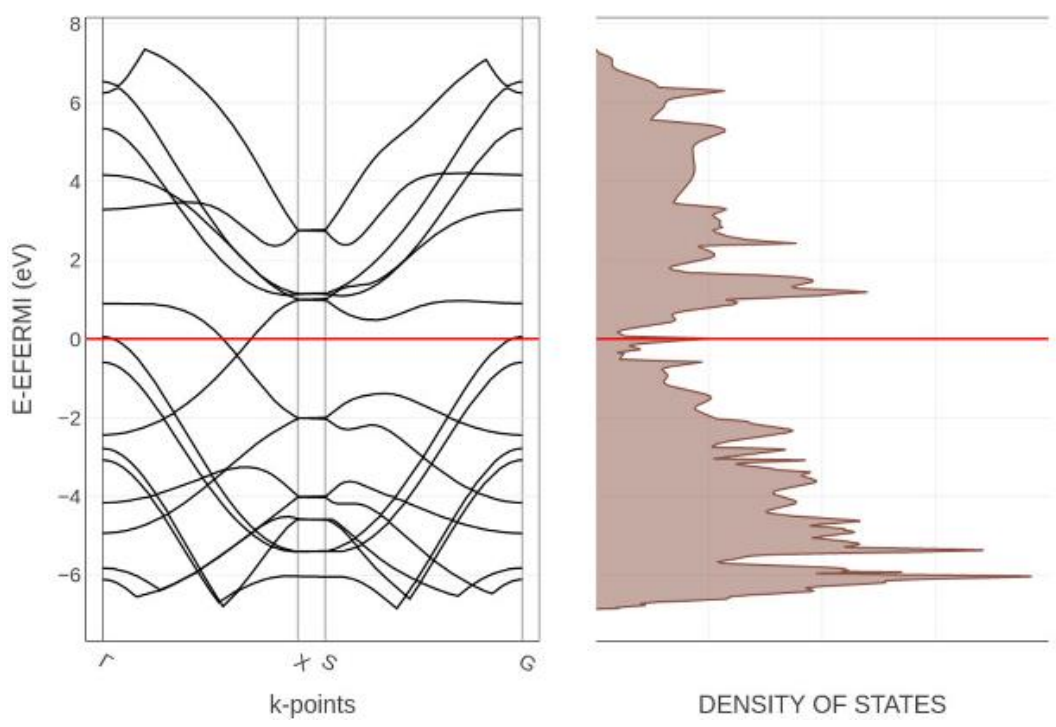<br>**and structures 6-h$^3$L, 6-h$^2$L, 6-hL, 6-hL$^2$, 6-hL$^3$** | **Structure 6-L**<br>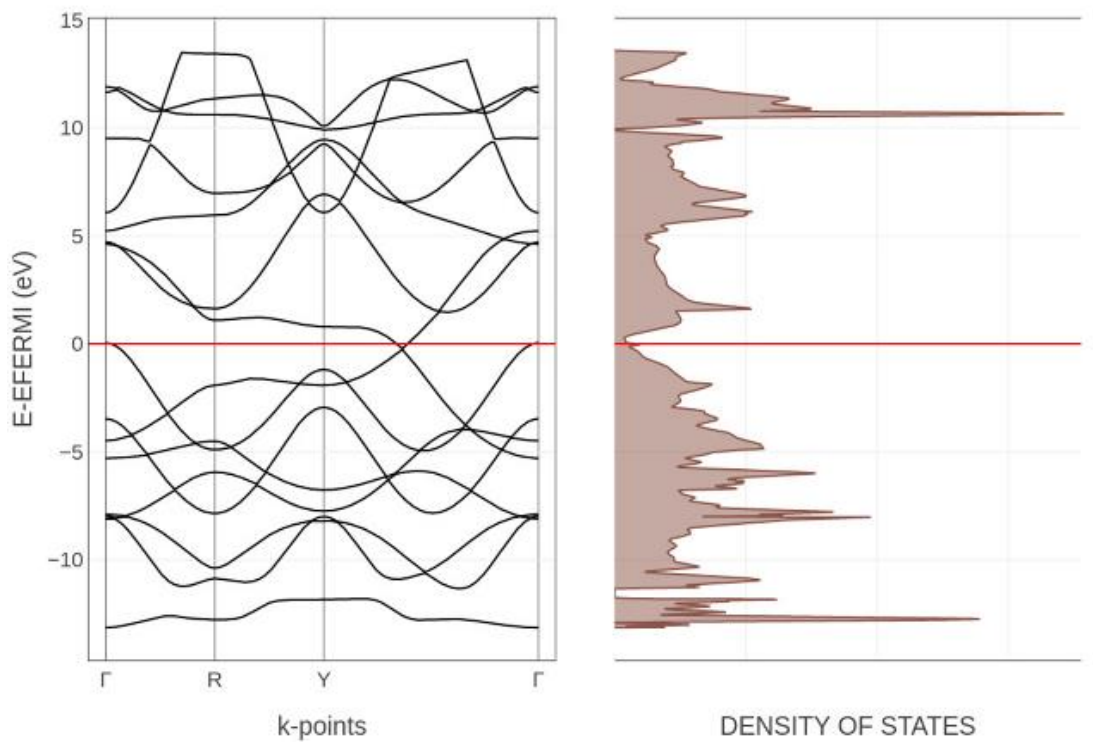 |
| **zero-gap semicond. - Dirac cones** (in addition to graphene and α-GDY) | **Structure 6-HR**<br>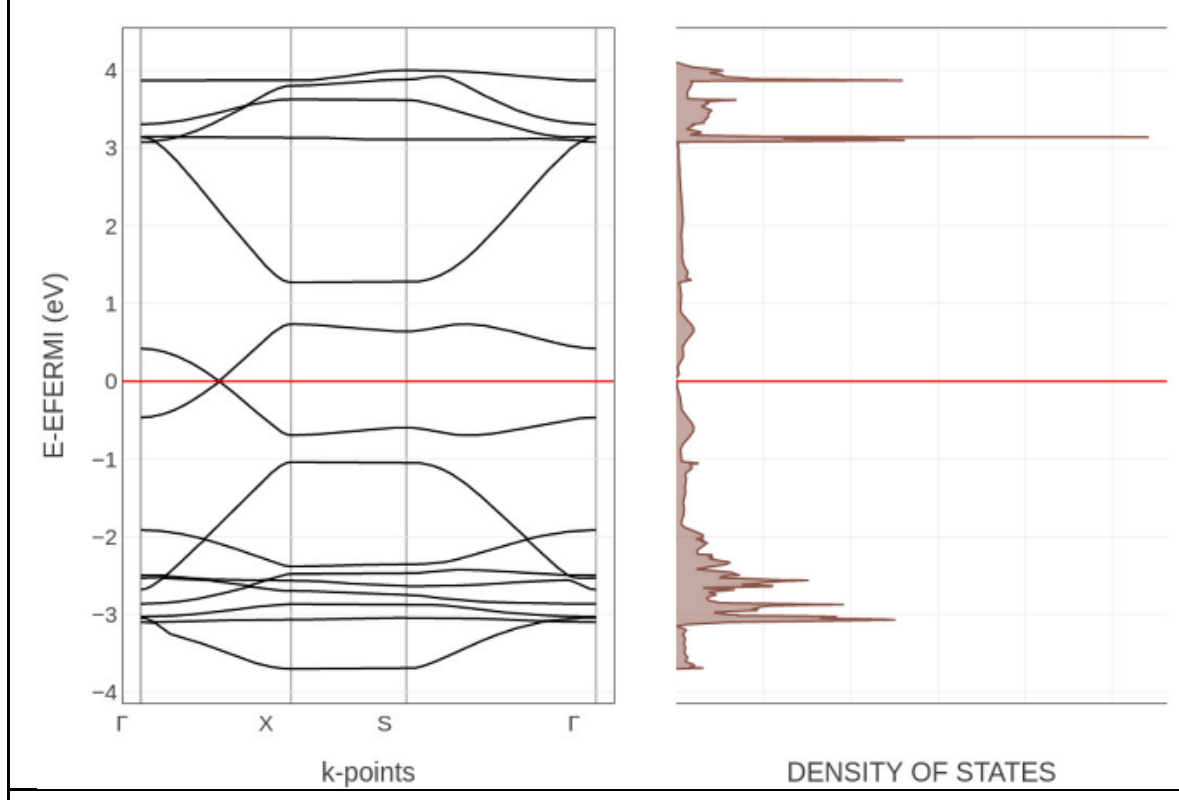 | **and structures 6-R, 6-H$^2$R$^2$, 6-HR$^3$, 6-H$^3$R, 6-hRL$^2$, 6-HL$^3$, 6-hR$^3$, 6-HR$^2$L, 6-HR$^2$ and 6-H$^2$R** |
| | **Structure 6-RL**<br>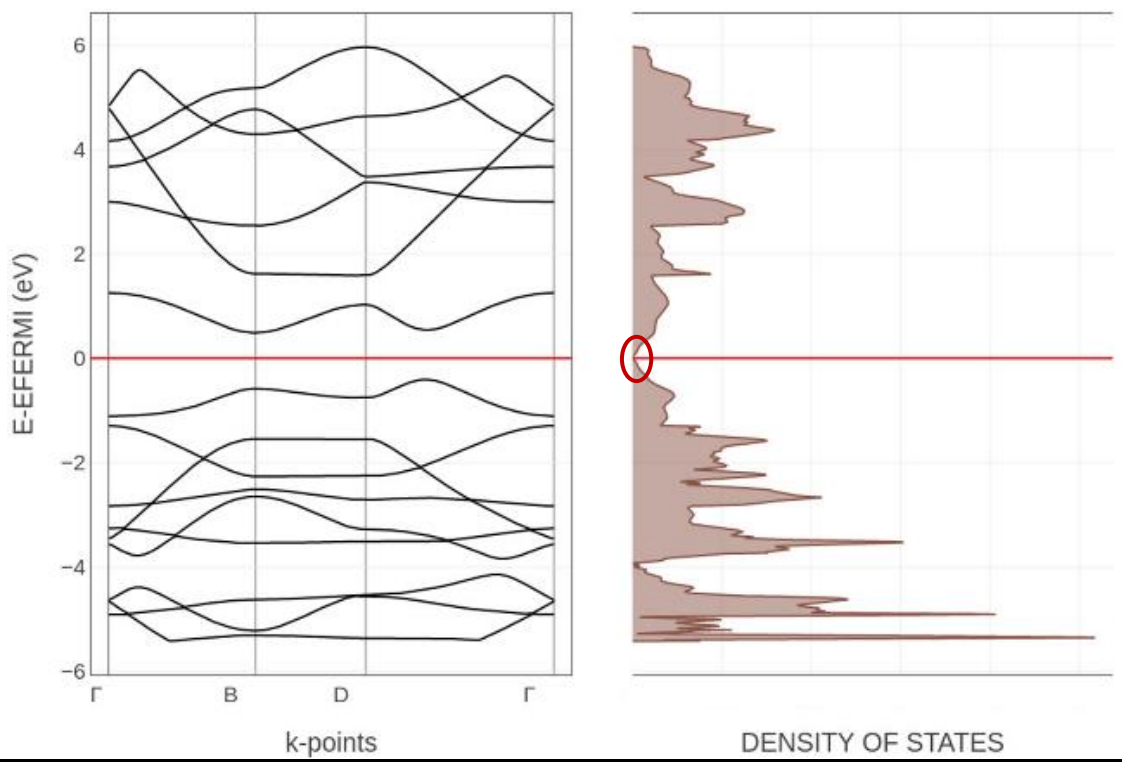 | **and structures 6-RL$^2$ and 6-R$^2$L**<br>In DOS, as circled in red, we find evidences for the existence of a zero-gap with Dirac cones |

***Figure 5:*** Comparison of DFT computed band structures and DOS of all the other structures here investigated.

The second class is formed by metallic systems and include structures made by graphene ribbons interconnected by diacetylenic groups (here called hL), also described as graziynes [49]. All these crystals (i.e. 6-h$^3$L, 6-h$^2$L, 6-hL, 6-h$^2$L$^2$, 6-hL$^2$, 6-hL$^3$, 6-L) share the same trend in the band structure. They present a half-filled band, in a structure characterized by a Dirac cone immediately below the Fermi Energy. In this class we find also structure 6-L, the limiting case of a 2D crystal formed by lines units, only.
The third class, collecting the largest number of different geometries, is characterized by zero-gap semiconductors presenting Dirac cones at the Fermi Energy. In this class, two subclasses with a slightly different behavior are presented: the first one, including structures 6-R, 6-HR, 6-H$^2$R$^2$, 6-HR$^3$, 6-H$^3$R, 6-

hRL$^2$, 6-HL$^3$, 6-hR$^3$, 6-HR$^2$L, 6-HR$^2$ and 6-H$^2$R, collects 2D crystals that, from the point of view of the band structure, present a pattern similar to α-GDY, with a Dirac Cone along one of the special directions in the BZ. The second subclass (RL) collects structures 6-RL, 6-RL$^2$ and 6-R$^2$L which formed already a cluster in relative energy: for these three crystals, Dirac cones cannot be identified along the three main path in the BZ but are located elsewhere in the Brillouin zone. The presence of Dirac cones is outlined by the linear behaviour of the DOS close to the Fermi energy. Even for these structures sharing a peculiar topology (they are crystals formed only by rhombus R and lines L units), 6-RL$^2$ and 6-R$^2$L are dual one respect to the other and 6-RL is self-dual.
To support the significance of these results, for some structures geometry optimization and band structure/DOS calculations have been repeated using HSE06 functional together with the same 6-31G(d) basis set, in order to check the effect related to the functional choice. As also demonstrated in our previous work on γ-GDY (6-hT$^2$), this functional can give indeed a more accurate quantitative evaluation of the band gap, in agreement with benchmark GW computed values, even if the qualitative trends are the same obtained with PBE0 [29]. The comparison between band structure/DOS computed with these two functionals are reported in the Supporting Information (Table S1). As expected, HSE06 band gap values are lower: 1.11 vs 1.63 eV for 6-hT$^2$ (γ-GDY), 0.34 vs 0.83 eV for 6-hT$^2$R and 0.68 vs 1.14 eV for 6-HT$^2$ (β-GDY) but still confirm that these structures are finite band gap semiconductor. A discrepancy is found however for 6-HT$^2$L structure, showing a very small (0.18 eV) but finite band gap with PBE0, while it is predicted as a 0-band gap semiconductor with HSE06. This result show that in presence of very small band gaps, the choice of the theoretical method could play an important role.
These trends reveal the topology-related shape of the band structure and the metallic, finite- or zero-gap semiconductor behaviour.

An inspection to the geometry of these structures allows further insight in how the building units (Fig. 1) determine the behaviour of the 2D carbon materials. A finite-band gap can be the consequence of the presence of T units, while hexagonal units (h and H) are present in zero-band gap semiconductors with Dirac Cones, as evidenced by the limiting case of graphene (h units, only) and α-GDY (H units, only). R units (see the case of 6-R) are similar to h and H, promoting the occurrence of zero-gap semiconductive behavior with Dirac cones, while L units, as in the case of 6-L (only L building blocks) occur in structures showing a metallic band structure. We can conclude that whenever h, H or R units are present there would be a tendency towards zero-gap semiconductors showing Dirac cones, when L units are present a tendency towards metallic structures while, on the other hand, T units are related to a gap opening between valence and conduction bands.
All metallic structures are formed by systems where only h and L units are present and the effect of L seems to dominate on that of h in affecting the band structure. Zero-gap structures all contains H and/or R and for some of them also h and L units can be present: in any case, H/R units seems to dominate the behavior of band structure. Finally, T units seem to dominate over all the other units in affecting the band structure, promoting a band gap. Interestingly, 6-HT$^2$L represents a limiting case showing a balance between the opposite effects of T and L units. The dominating effect seems to be related to the functional choice since we find 0.18 eV for PBE0 and zero-gap for HSE06.
For the systems already reported in literature our results are consistent and describe the same behaviour. [17]. However,the band gap values are lower than the ones here reported for 6-HT$^2$ and 6-hT$^2$, probably due to the use of a pure GGA functional (PBE) with respect to the hybrid one used in our calculation. A discrepancy is found for 6-hT$^2$R structure, for which we predict a gap of 0.83 eV with PBE0 (and 0.34 eV with HSE06) while a zero-gap semiconductor is predicted by Park et al. [17]. This point out again that for semiconductive systems having a small band gap (i.e. below 1 eV), the choice of the functional could be relevant to predict the behavior of the materials in terms of its electronic structure. Based on the role of the different h, H, R, L and T units and considering the results for 6-$HT^2L$ with PBE0 and HSE06, the choice of the functional is relevant in predicting the lower or larger dominating effect of the different units in affecting the band structure behaviour. This is not a primary effect when large band gaps are present, but some peculiar cases could require more attention to the theoretical method.

## 4. Conclusions

The possibility to develop sp-$sp^2$ carbon 2D materials by playing on the topology and connectivity of sp and $sp^2$ domains or on their relative ratio is clearly appealing for both fundamental and applied research with possible outcomes in technology. Up to now, most of the experimental research on GDY systems focused on the development of proper synthesis techniques while a huge amount of theoretical investigation has been focused on the properties of these materials. However, most of the literature focused on the $\gamma$ polymorph of GDY and related systems (nanoribbons, molecular fragments…) or in some cases on the $\alpha$- and $\beta$-GDY. Only a few other possible structures have been considered so far. There is ideally a wide range of possible sp-$sp^2$ carbon materials, in form of 2D crystals, that could be possible alternatives to $\gamma$-GDY.

We proposed a computational investigation aimed at the molecular design of new sp-$sp^2$ carbon 2D crystals, focusing in particular on the importance of the structure and topology in modulating the relative energies and the band structure with respect to graphene. Our approach is able to predict all the possible sp-sp2 crystals as graphene derivatives. By restricting our search to all the sp-$sp^2$ carbon crystals with a maximum number of eight $sp^2$ carbon atoms per unit cell, we generated 26 2D crystals. Periodic boundary conditions DFT simulations have been carried out revealing some peculiar trends both in relative energy and electronic properties, which can be described in terms of general topological effects.

In all the cases an increase of the sp-$sp^2$ carbon ratio produced an increase of relative energies with respect to graphene, with two peculiar trends. A first one is constituted by graphene stripes interconnected by diacetylenic bridges (graziynes) which have been also predicted to a have a metallic behavior. The second trend collects 2D crystals (including $\alpha$-, $\beta$- and $\gamma$-GDY) which can be described in terms of common geometrical units formed by the carbon atoms, including h and H hexagons, L lines, R rhombus and T triangles. Describing the crystals in terms of these units allowed us to rationalize both relative energies and band structure: the higher is the similarity to graphene units (i.e. h) the lower are the relative energies; on the other hand H units increase the relative energies up to the limiting case of $\alpha$-GDY and are characteristic of zero-gap semiconductors with Dirac cones.

These different units can play a role in determining the electronic behaviour of the material: triangular T units are indeed the structural/topological factor which promotes semiconductive materials with a finite band gap; on the other hand L units would promote metallic structures while h, H and R units tends to induce zero-gap semiconductors with Dirac cones. As a general rule, in structures formed by different units, L is found to have a larger effect than h, while H and R dominate over L and finally T seems to dominate all over the other, even if these relative effects have been shown to have a non-negligible dependence on the DFT functional choice in particular for some peculiar structures.

These findings give a relevant indication to develop properly new semiconductive sp-$sp^2$ carbon materials, since they are able to give general and simple topological rules to design new systems built as proper combination of simple building blocks (h, H, R, L and T units), where the electronic properties are properly tailored by precisely controlling their topology. This will offer many outcomes in view of applications and some insight for engineering new carbon nanostructured materials with tailored properties.

**Acknowledgements**

Authors acknowledge funding from the European Research Council (ERC) under the European Union's Horizon 2020 research and innovation program ERC—Consolidator Grant (ERC CoG 2016 EspLORE grant agreement No. 724610, website: www.esplore.polimi.it)

# Supporting Informations

## Designing all graphdiyne materials as graphene derivatives: topologically driven modulation of electronic properties

Patrick Serafini[1], Alberto Milani[1*], Davide M. Proserpio[2,3], Carlo S. Casari[1*]

[1]*Dipartimento di Energia, Politecnico di Milano, via Ponzio 34/3, Milano, Italy*
[2]*Dipartimento di Chimica, Università degli Studi di Milano, 20133 Milano, Italy*
[3]*Samara Center for Theoretical Materials Science (SCTMS), Samara State Technical University, Samara 443100, Russia*

*Corresponding authors: alberto.milani@polimi.it; carlo.casari@polimi.it

**Table S1:** Computed band structures and density of states (DOS), with a sketch of the structure for each of the analysed 2D crystal, are reported. For some specific cases, a comparison between band structures and DOS calculated with PBE0 and HSE06 exchange-correlation density functionals is reported.

| Names | Band Structure + DOS (PBE0) | Band Structure + DOS (HSE06) |
| --- | --- | --- |
| Graphene | 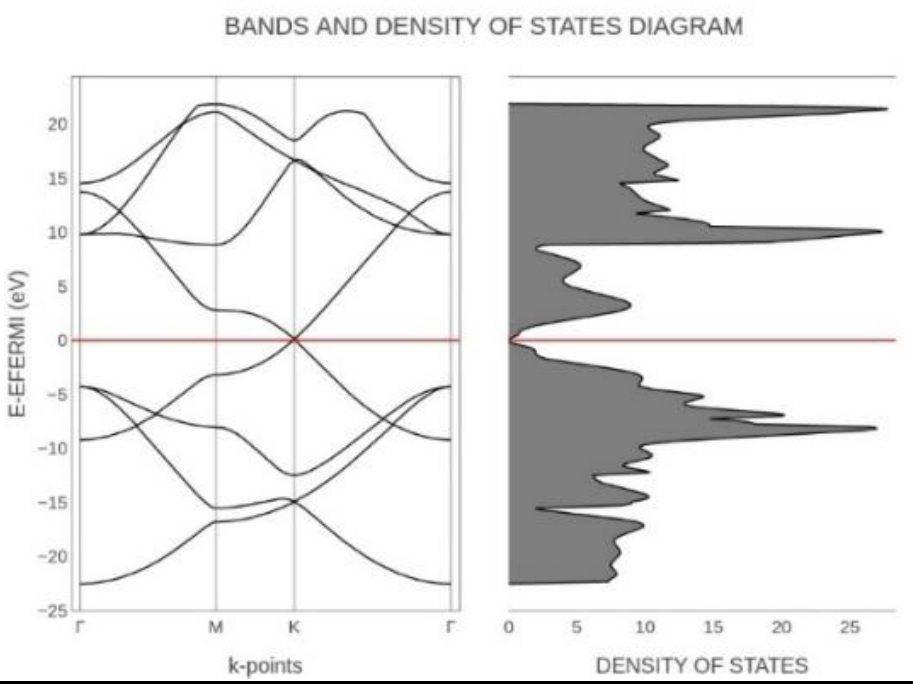  | |

<table>
<tr>
<td>6-L<br>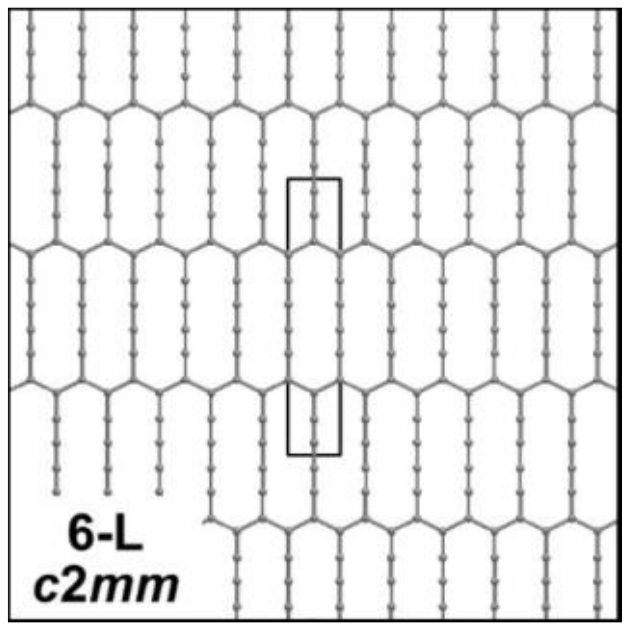
</td>
<td>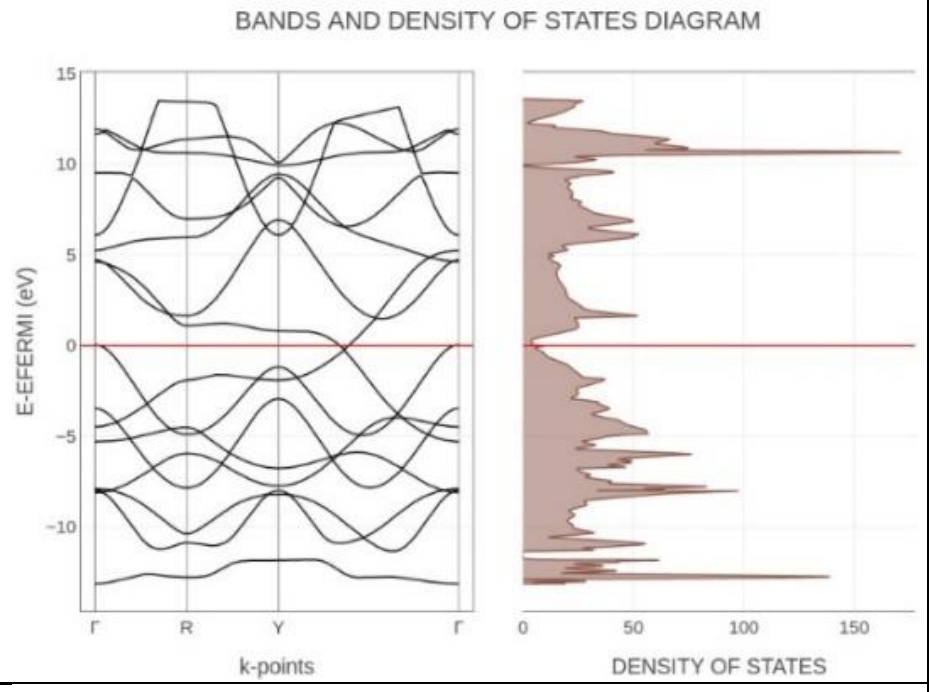
</td>
<td>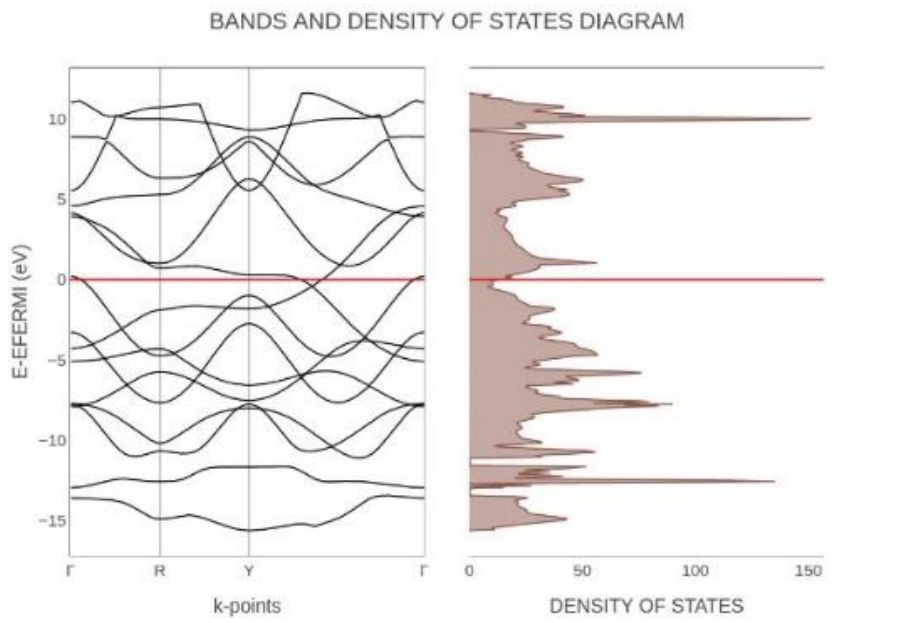
</td>
</tr>
<tr>
<td>6-R<br>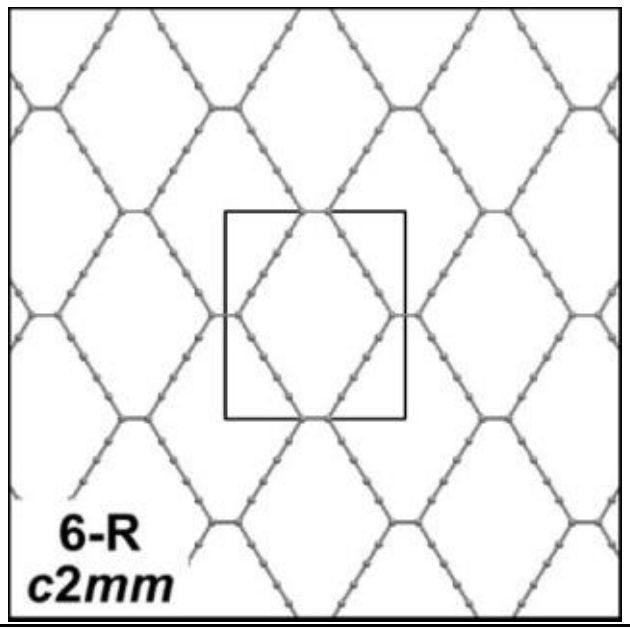
</td>
<td>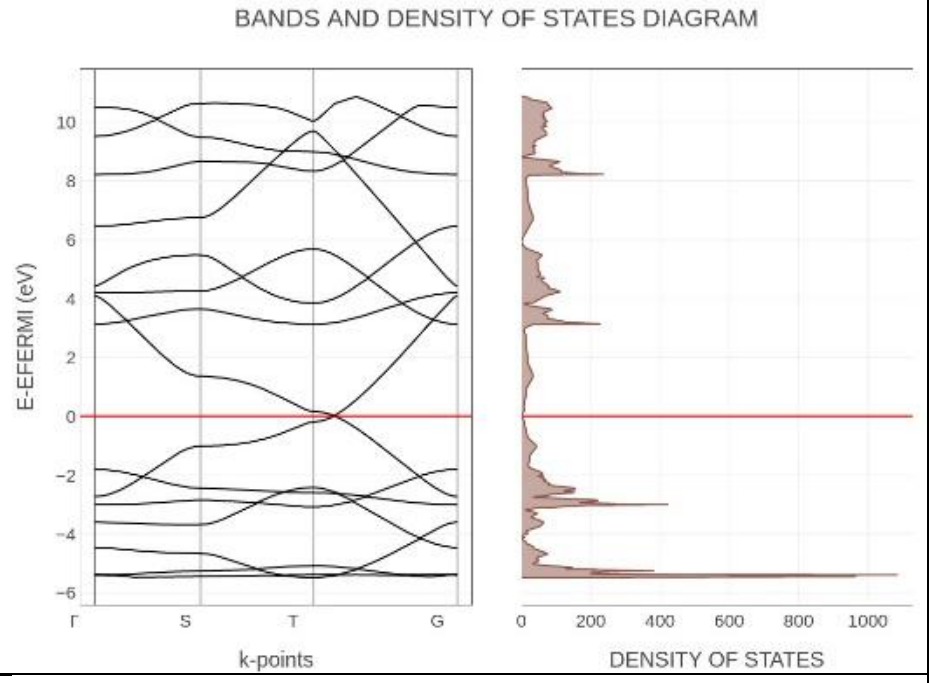
</td>
<td>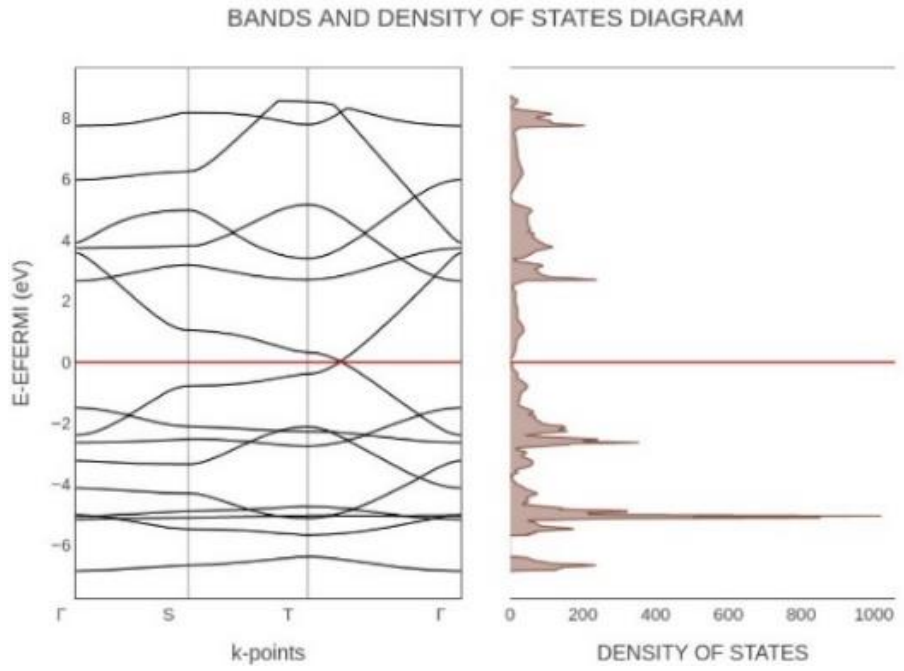
</td>
</tr>
<tr>
<td>6-hL<br>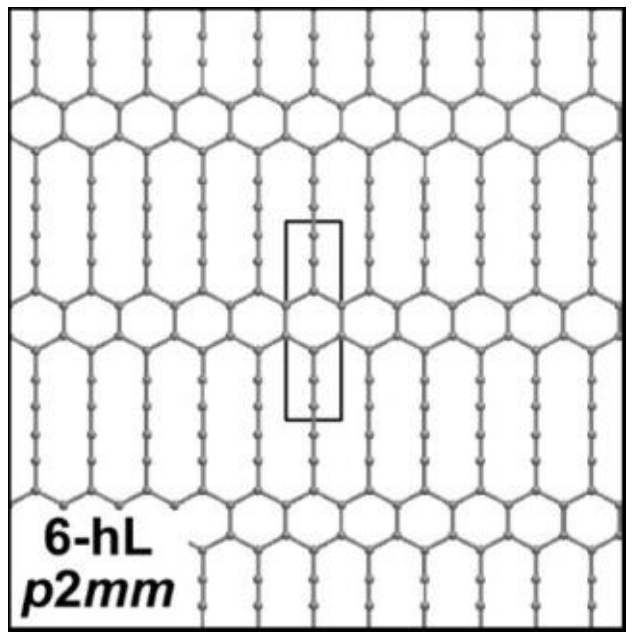
</td>
<td>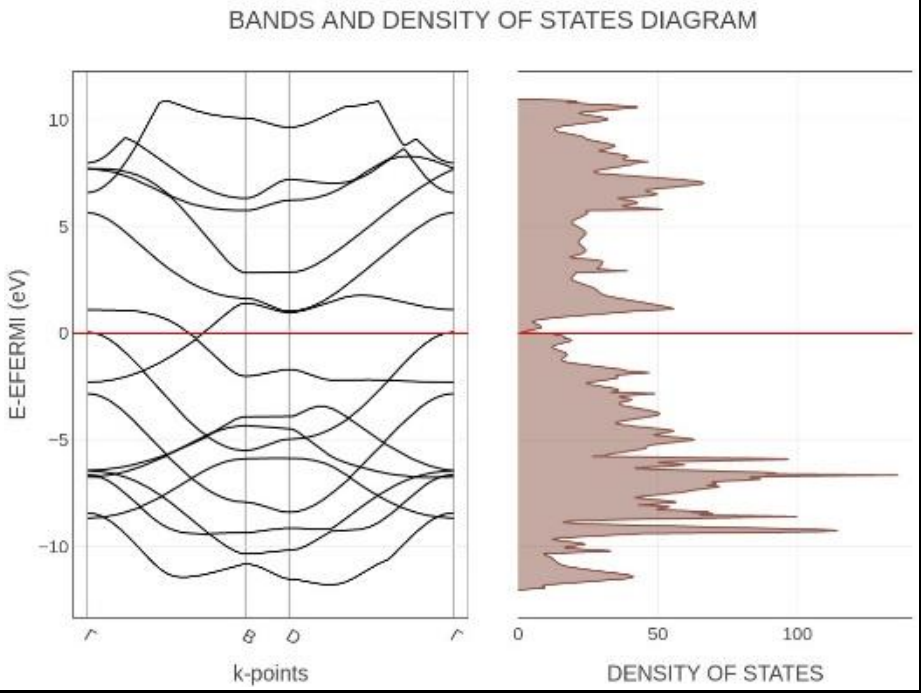
</td>
<td></td>
</tr>
</table>

| | | |
|---|---|---|
| 6-HR<br>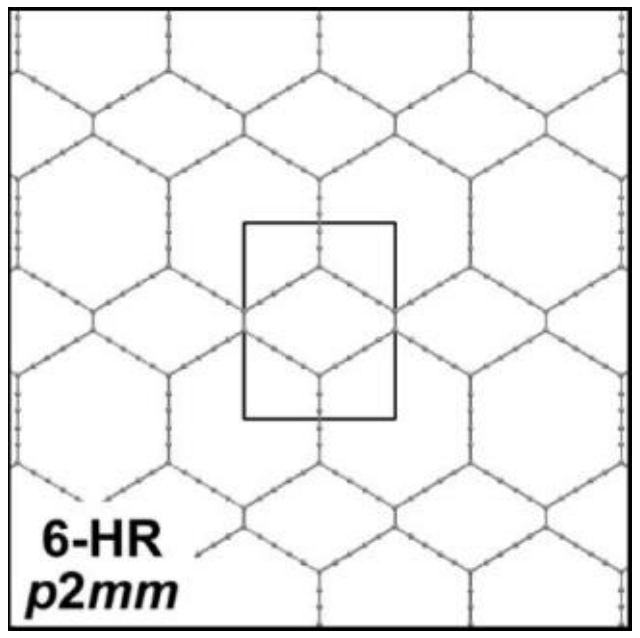 | 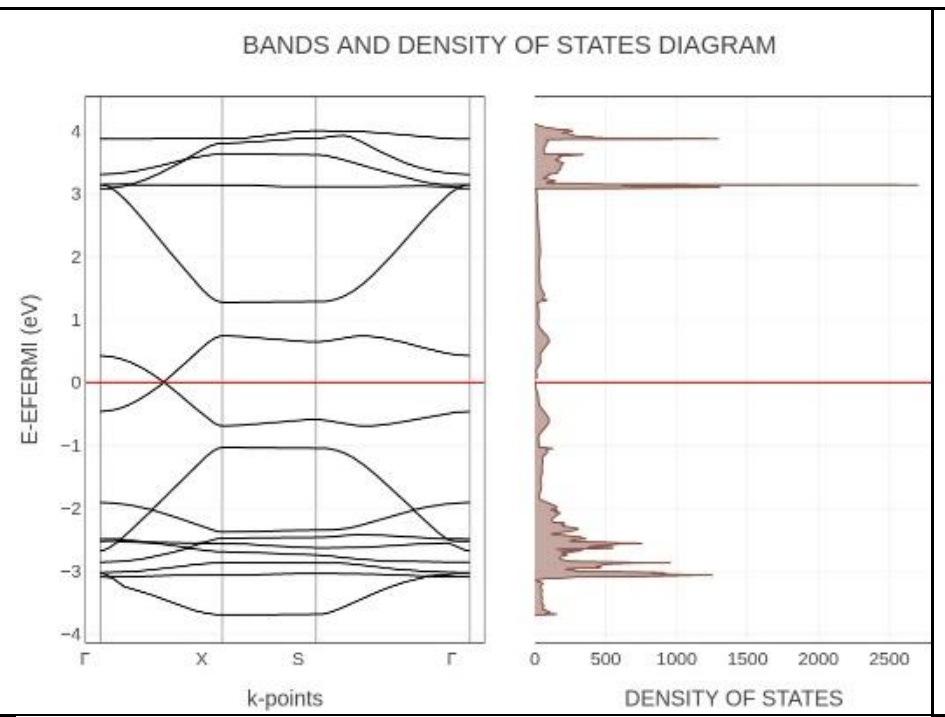 | |
| 6-h$^2$L$^2$<br>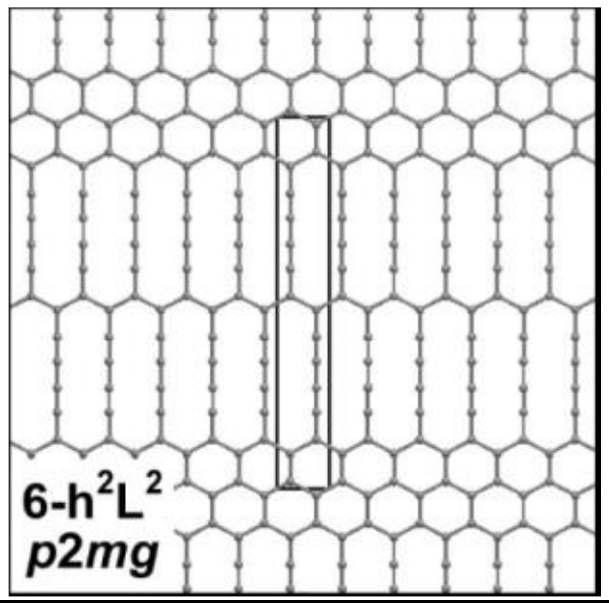 | 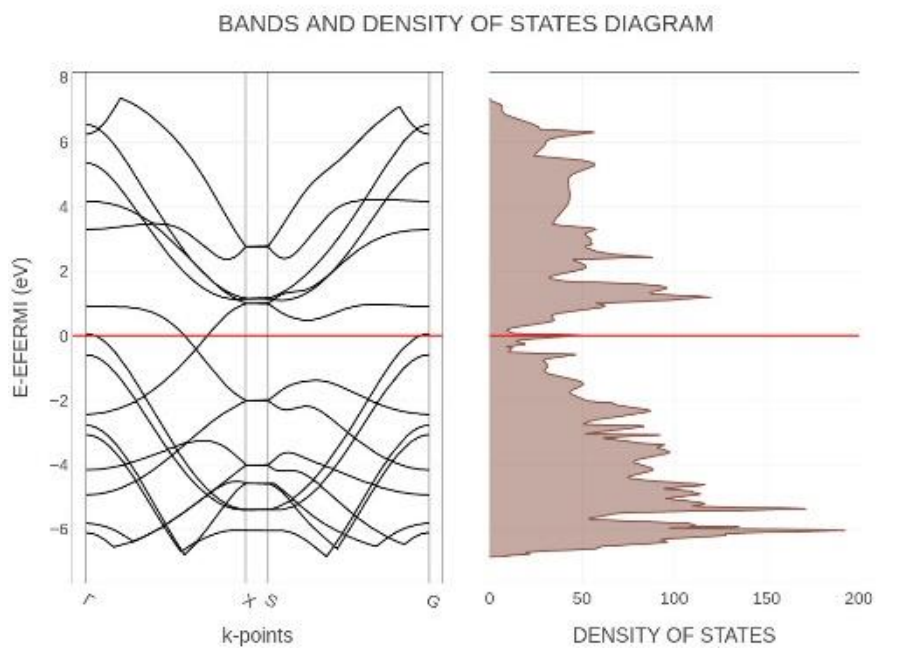 | 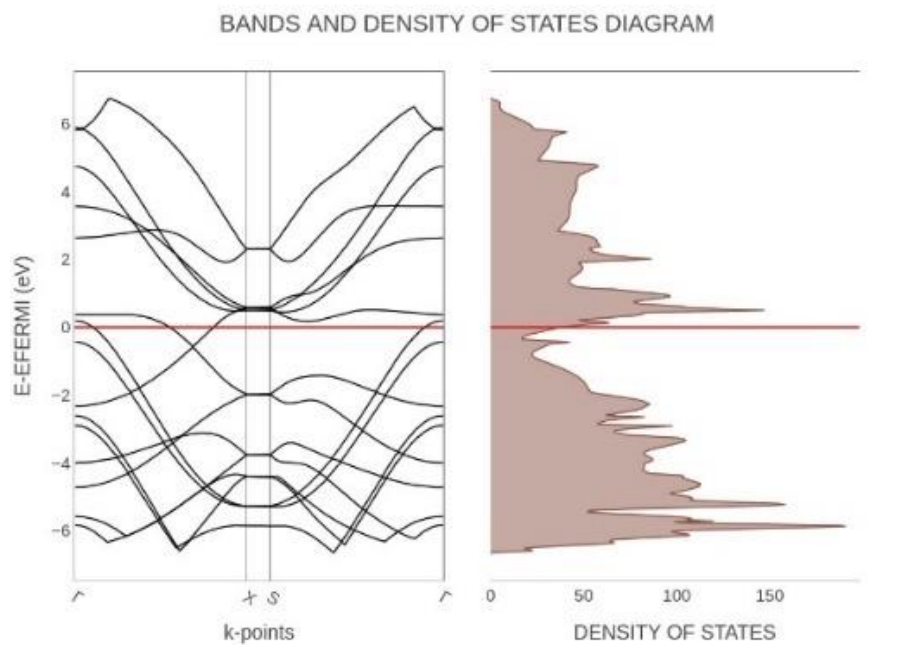 |
| 6-H$^2$R$^2$<br>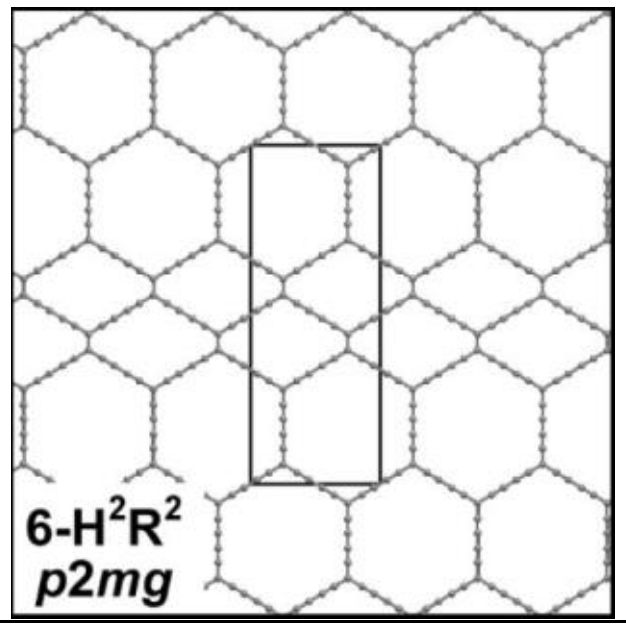 | 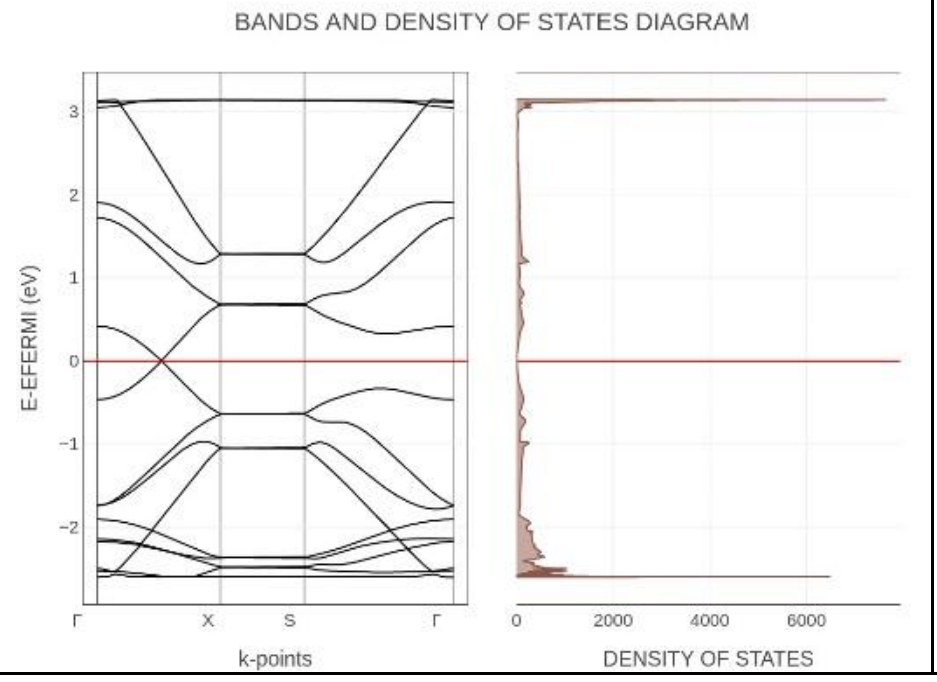 | |
| 6-h$T^2$R<br>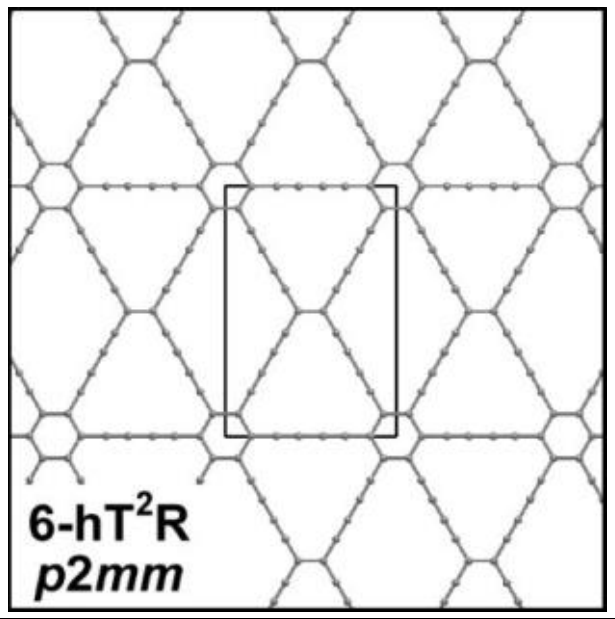 | 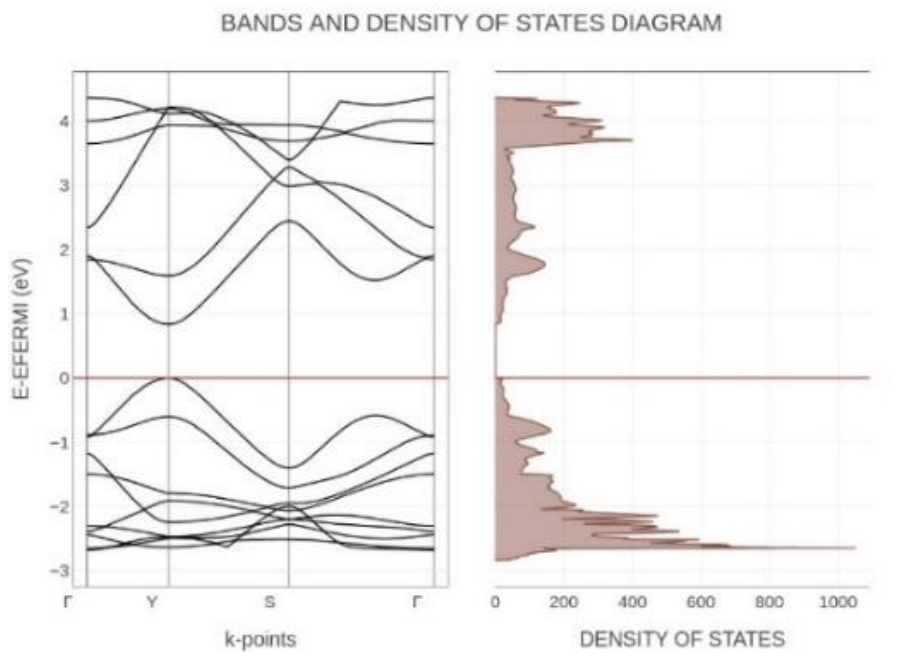 | 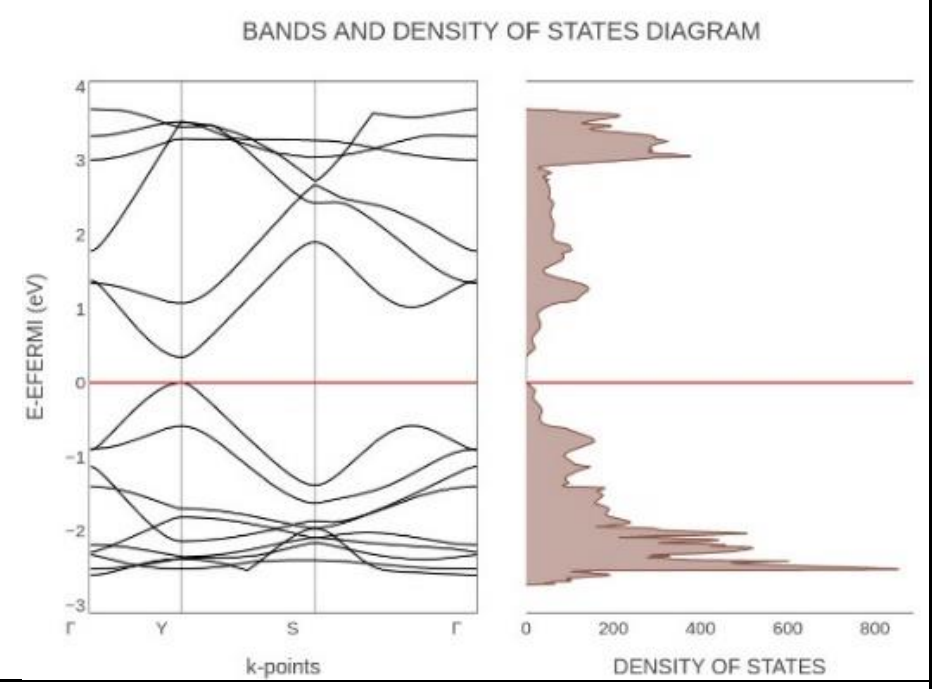 |
| 6-H$T^2$L<br>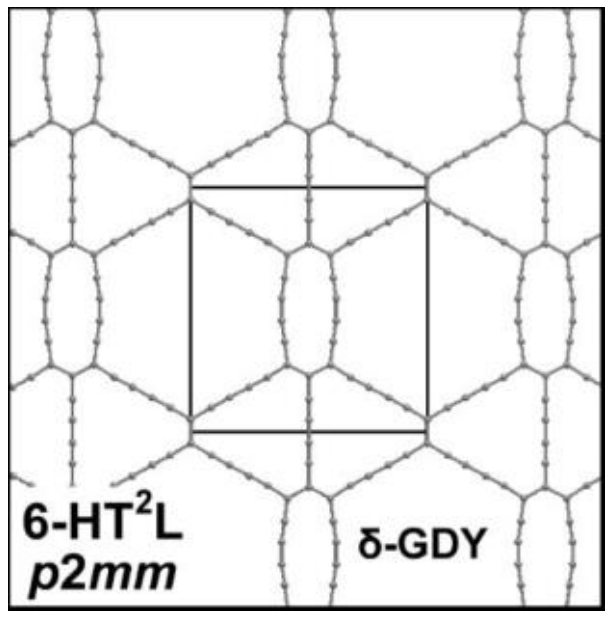 | 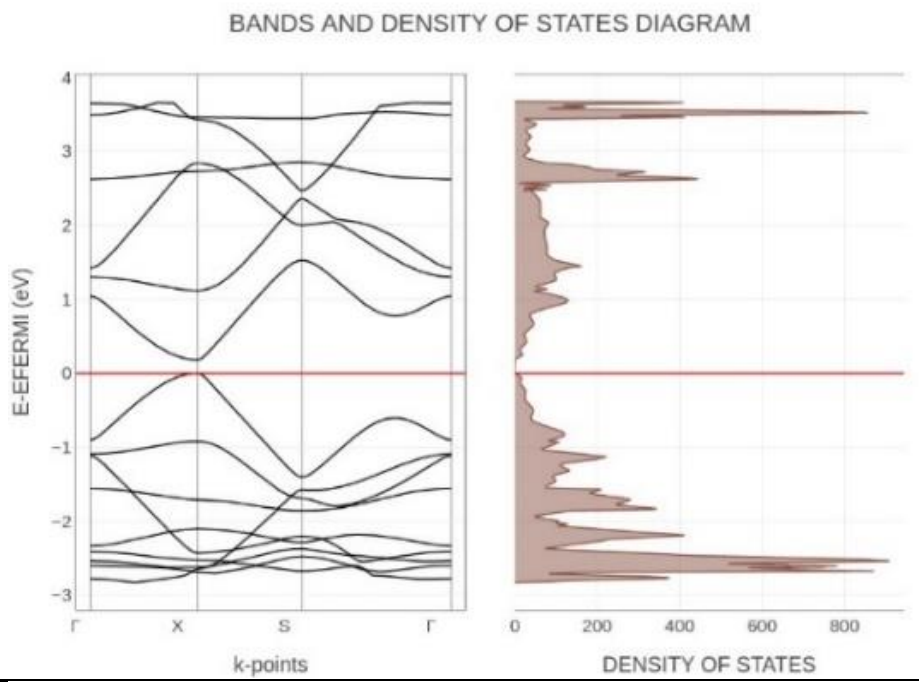 | 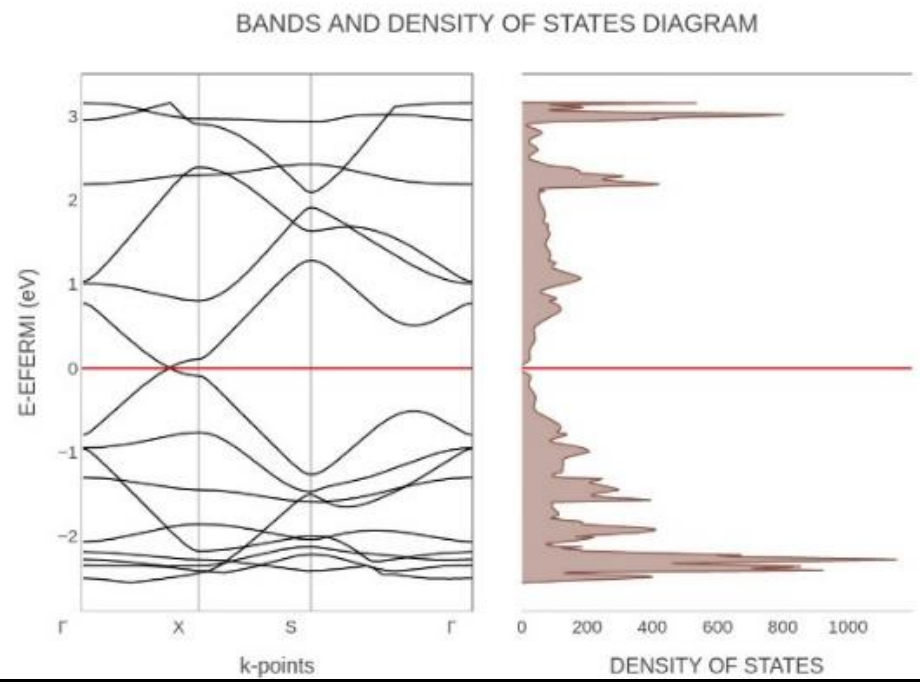 |
| 6-$h^3$L<br>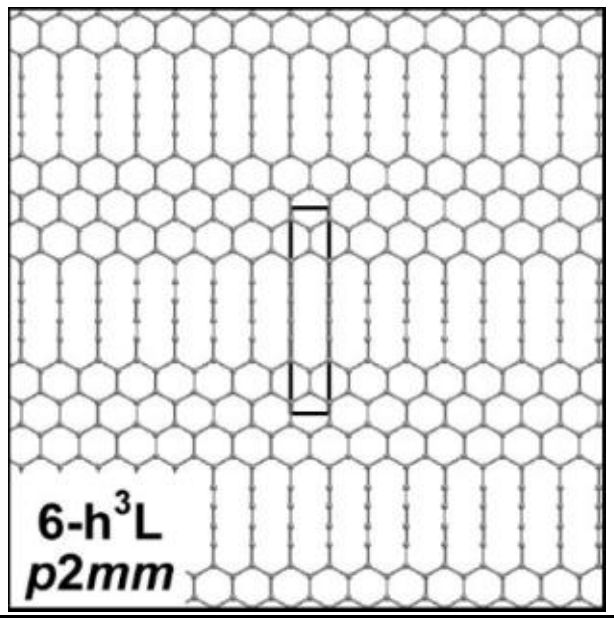 | 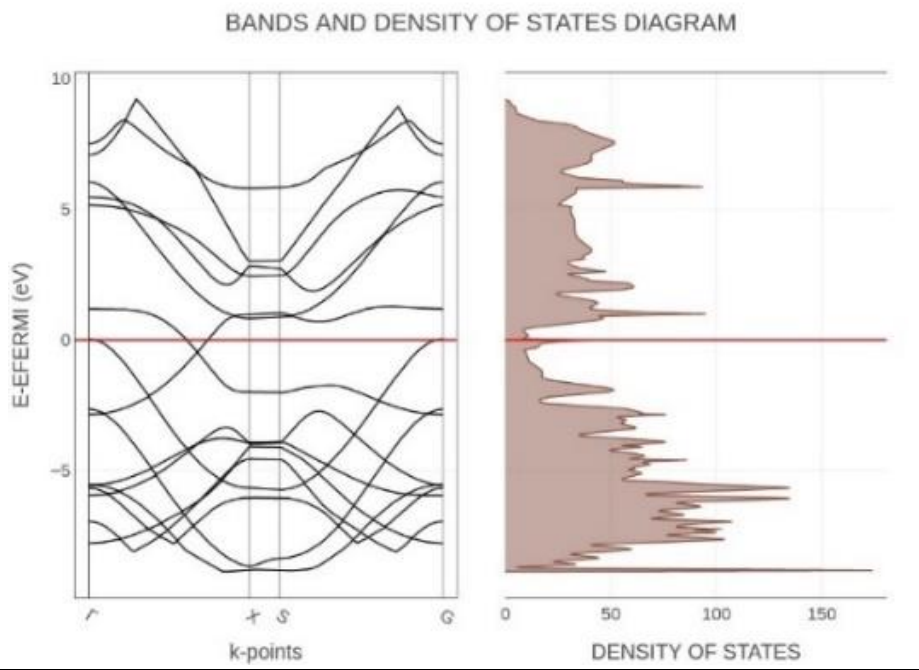 | |

<table>
<tr><td>6-hL$^3$<br>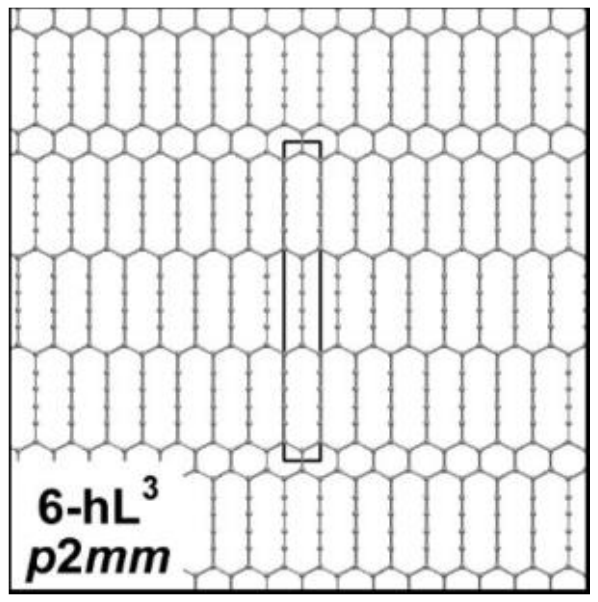</td><td>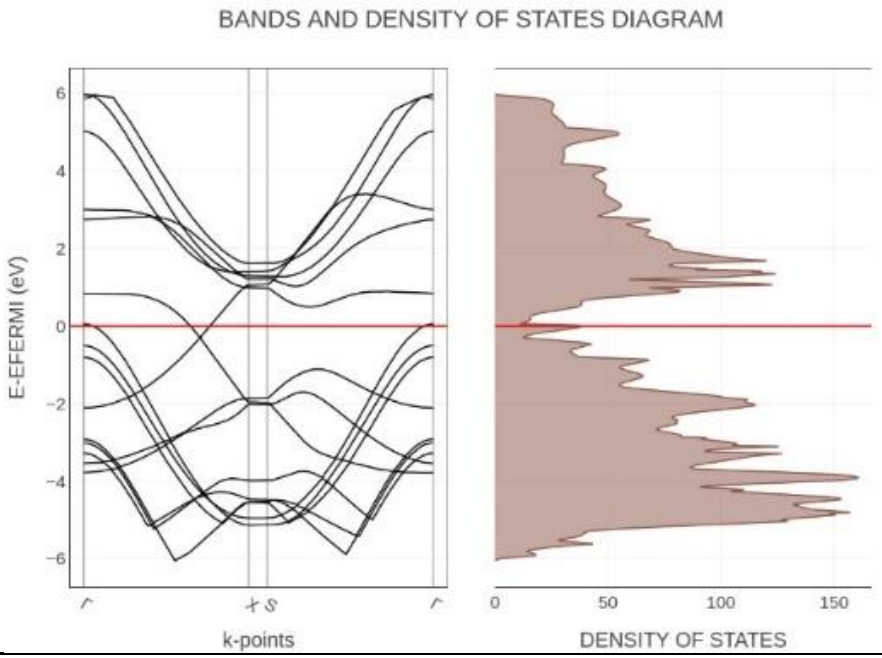</td><td></td></tr>
<tr><td>6-HR$^3$<br>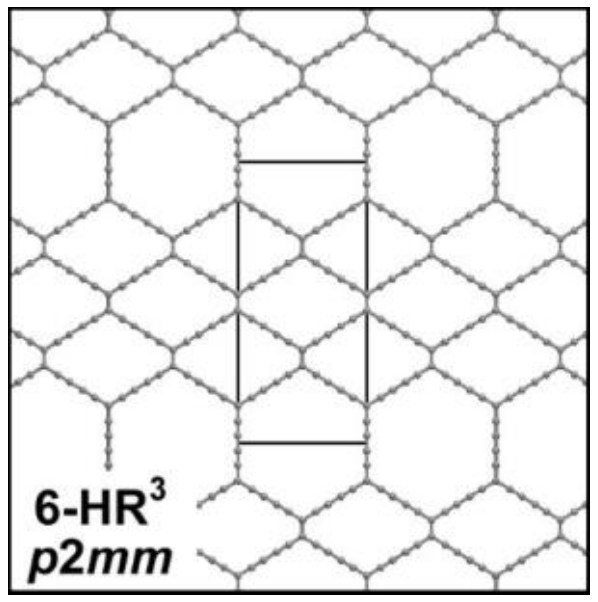</td><td>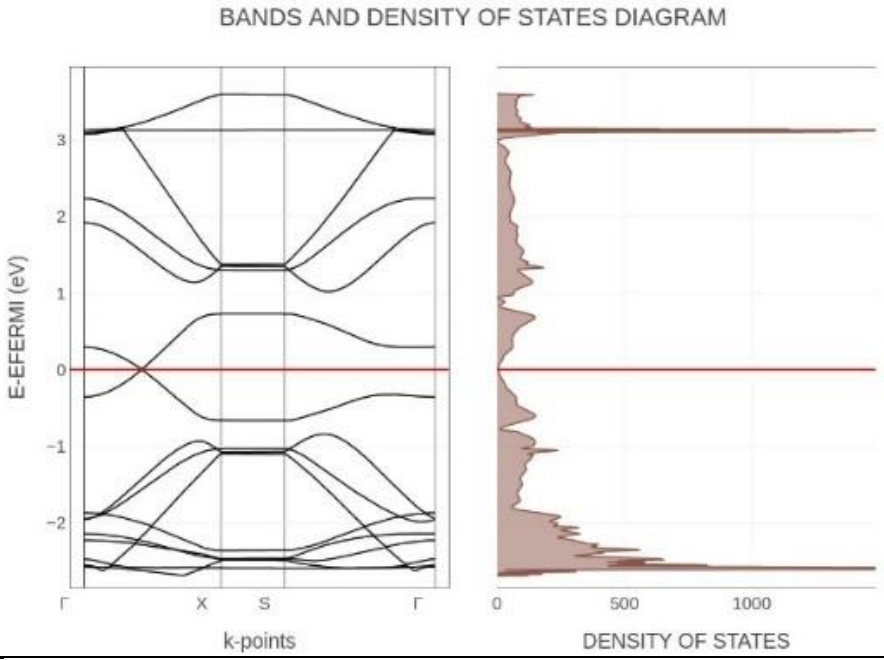</td><td></td></tr>
<tr><td>6-H$^3$R<br>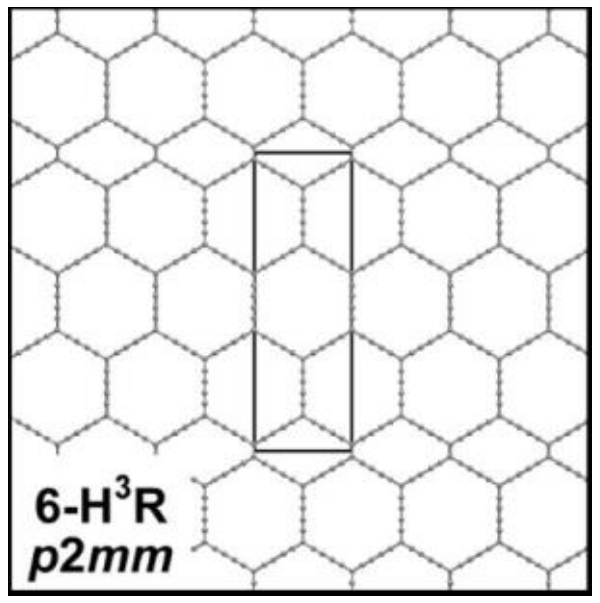</td><td>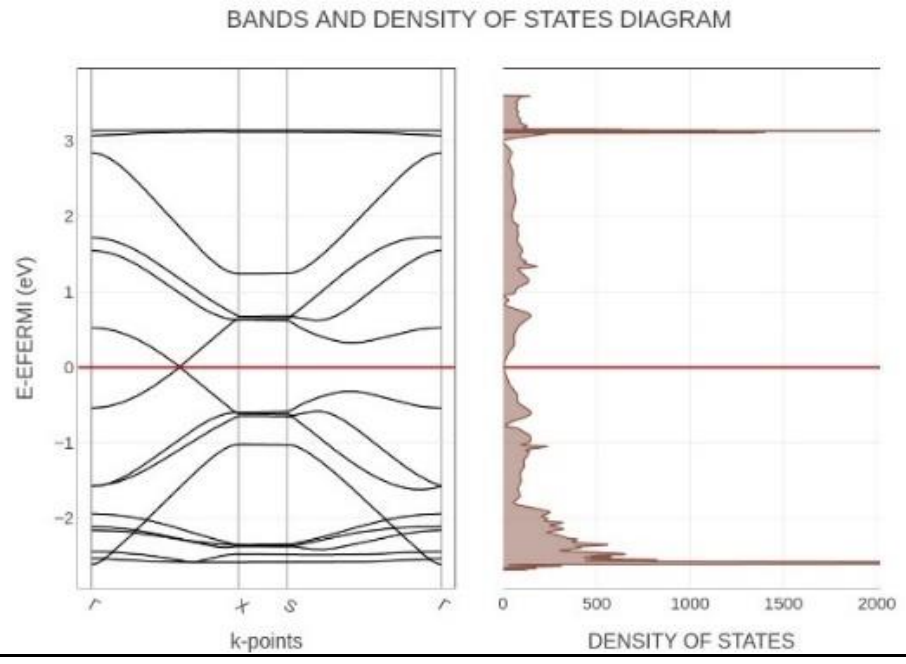</td><td></td></tr>
</table>

<table>
<tr>
<td>6-hRL$^2$<br>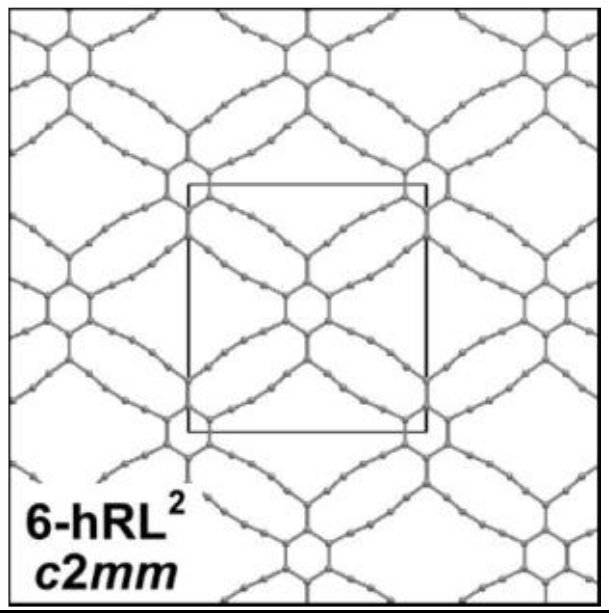</td>
<td>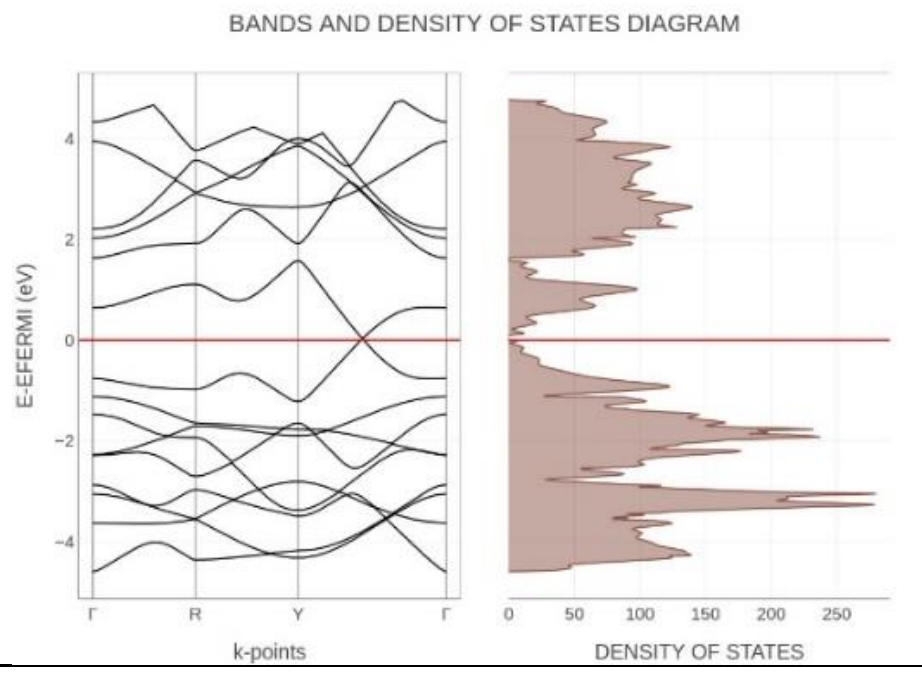</td>
<td></td>
</tr>
<tr>
<td>6-HL$^3$<br>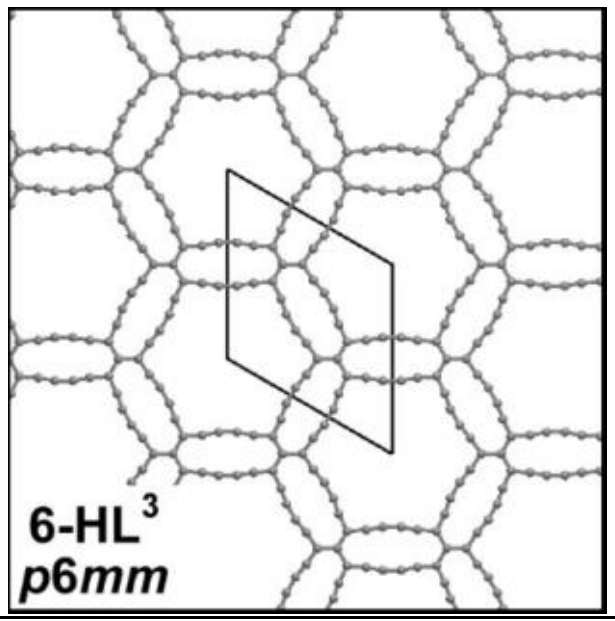</td>
<td>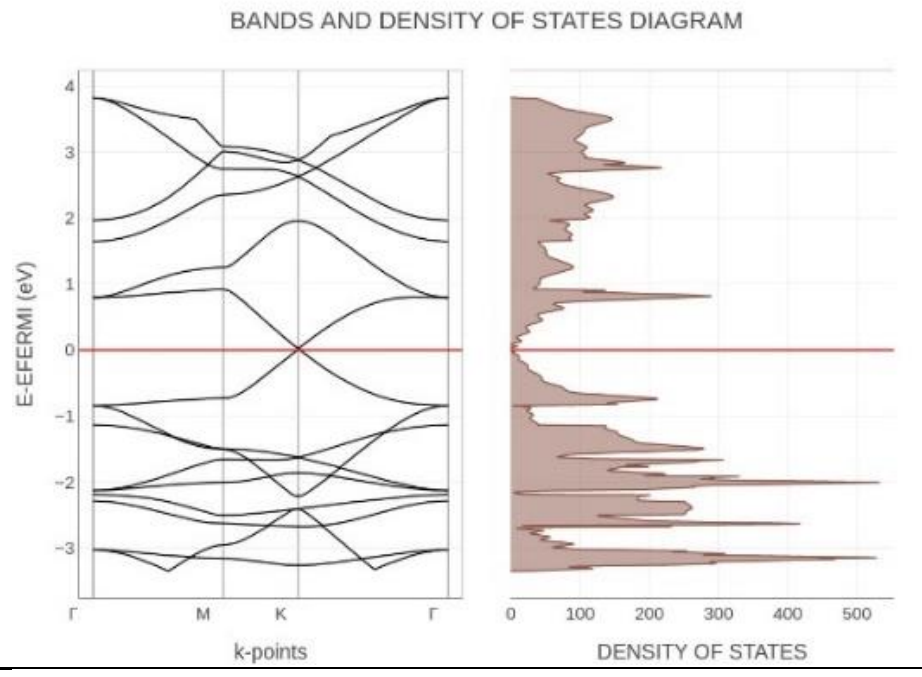</td>
<td></td>
</tr>
<tr>
<td>6-hR$^3$<br>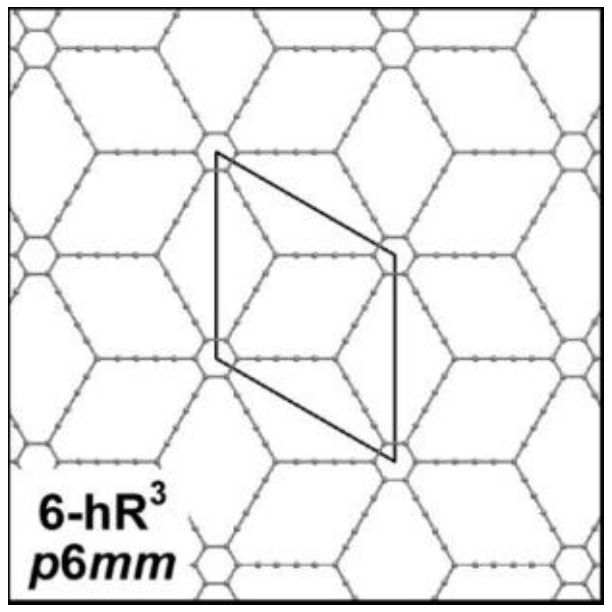</td>
<td>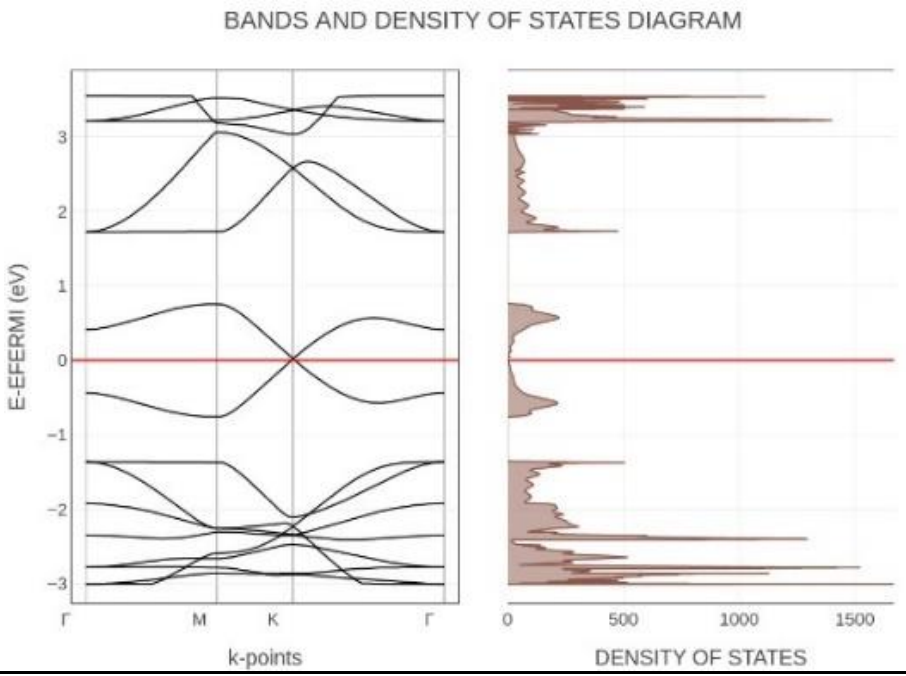</td>
<td></td>
</tr>
</table>

| | | |
|---|---|---|
| 6-HR$^2$L 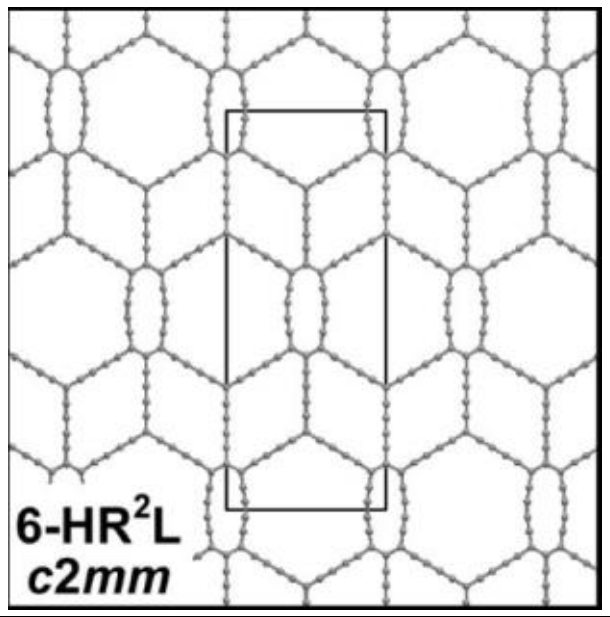  | 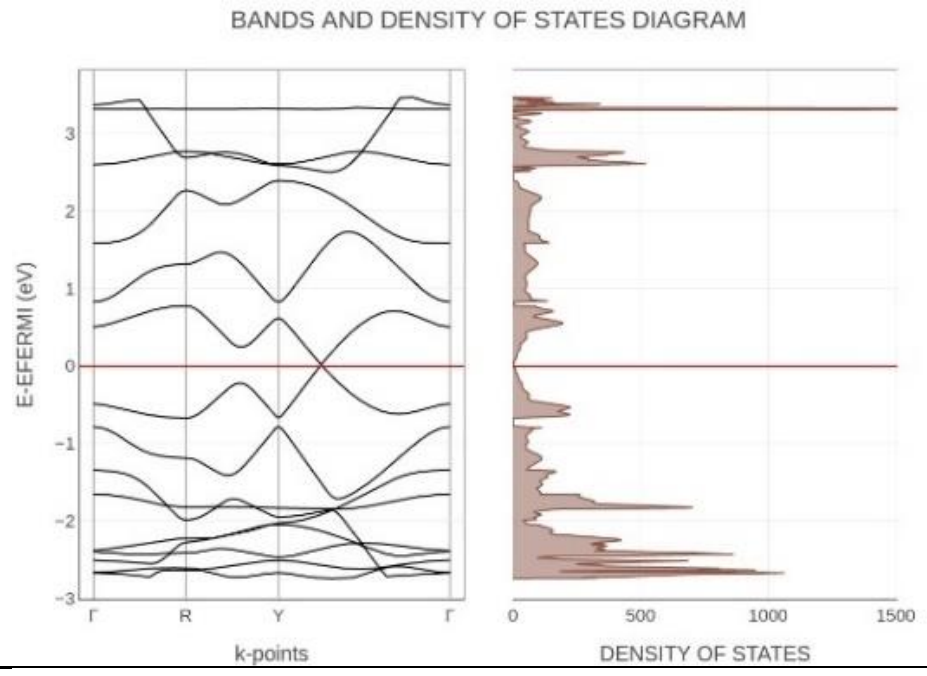  | |
| 6-h$^2$L 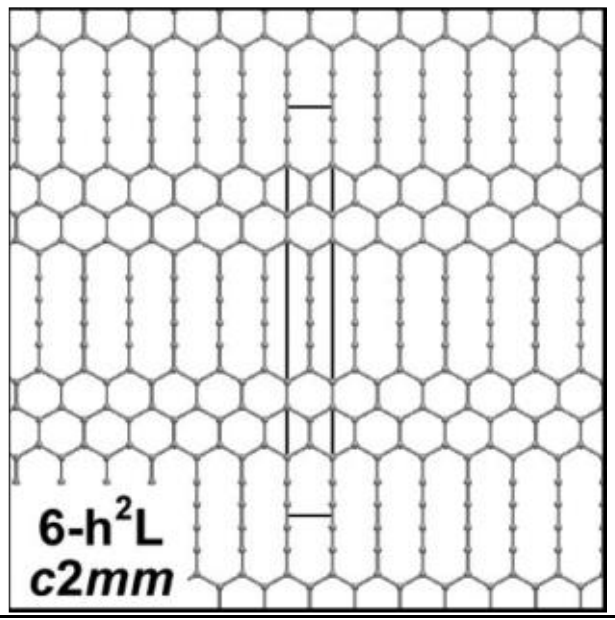  | 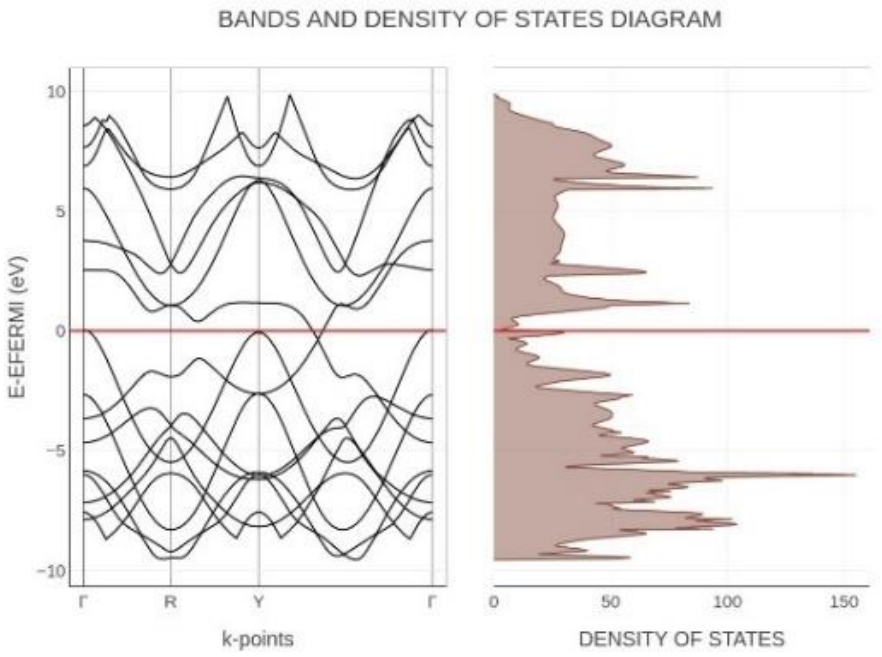  | |
| 6-hL$^2$ 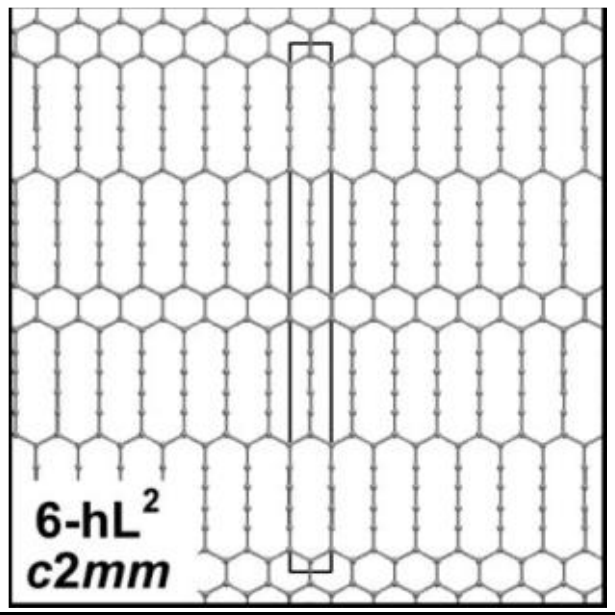  | 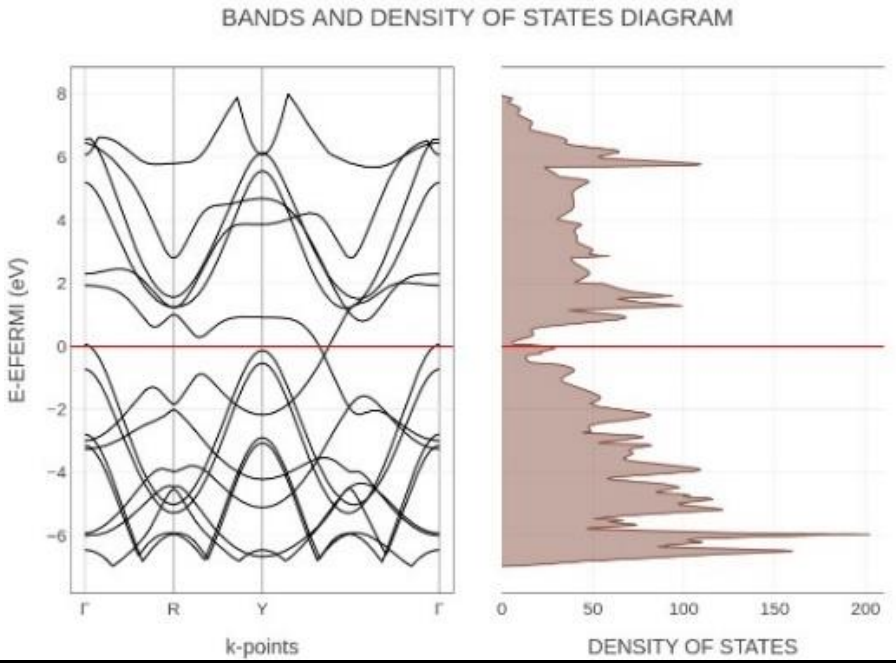  | |

<table>
<tr>
<td>6-HR$^2$<br>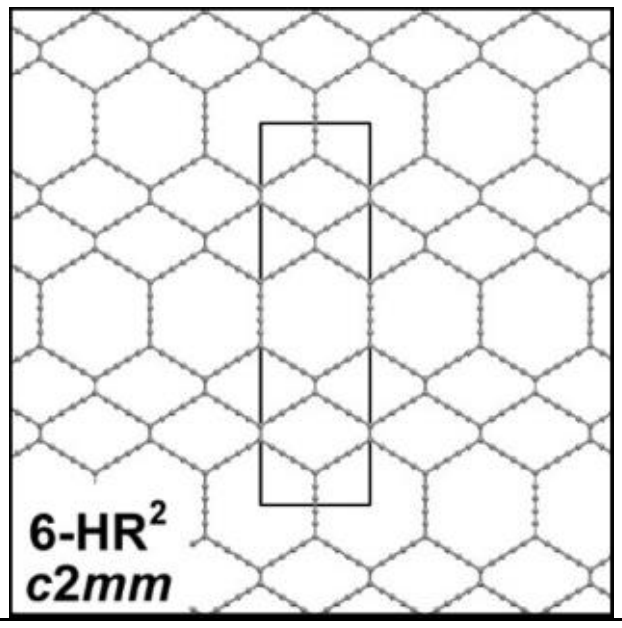</td>
<td>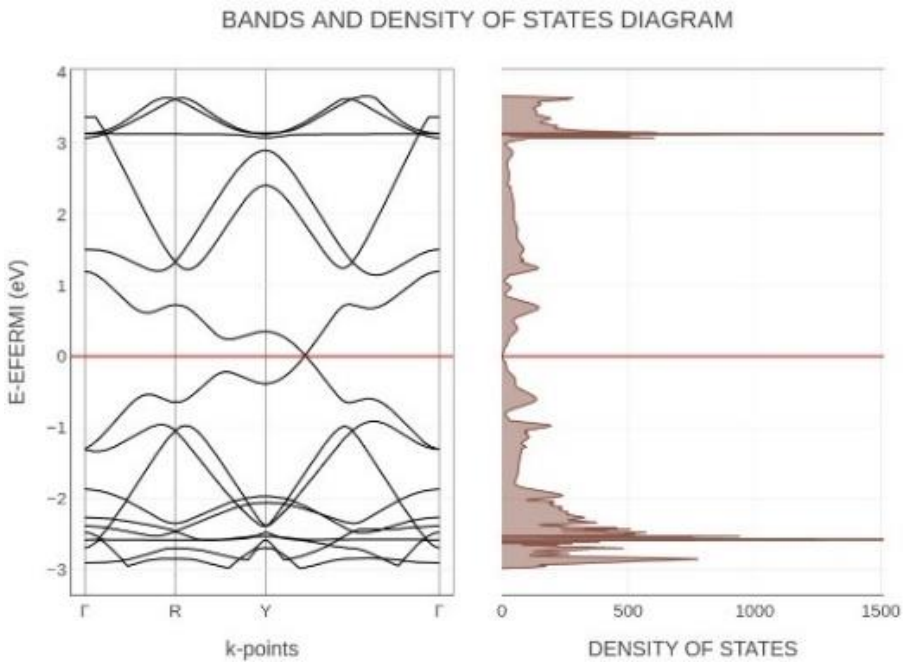</td>
<td></td>
</tr>
<tr>
<td>6-H$^2$R<br>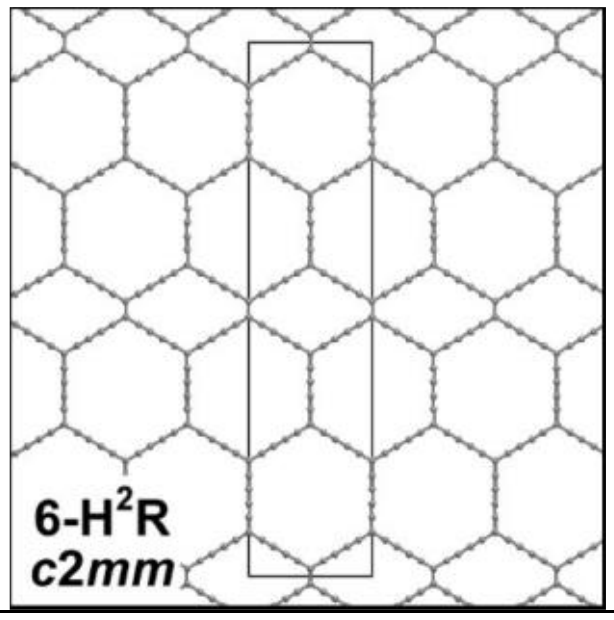</td>
<td>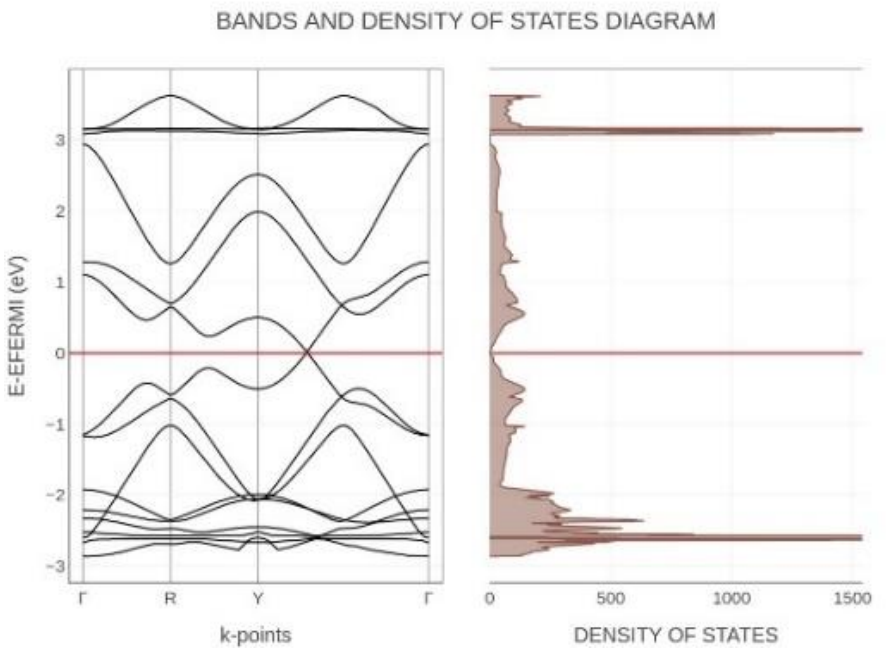</td>
<td></td>
</tr>
<tr>
<td>6-hT$^2$ (γ-GDY)<br>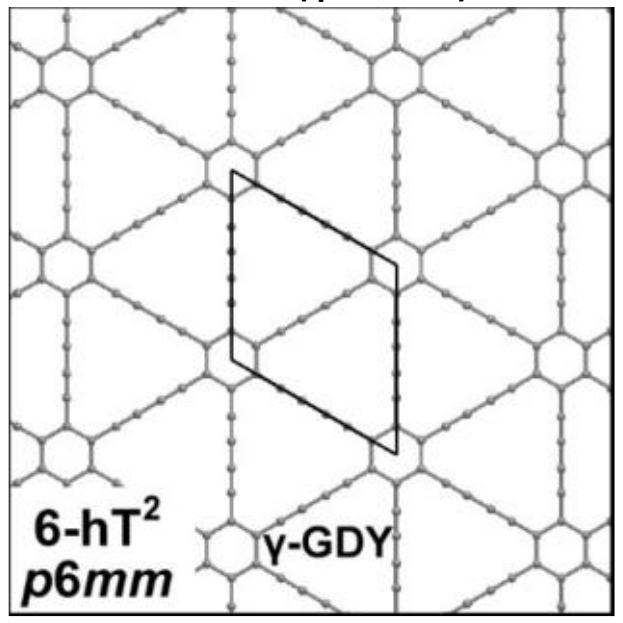</td>
<td>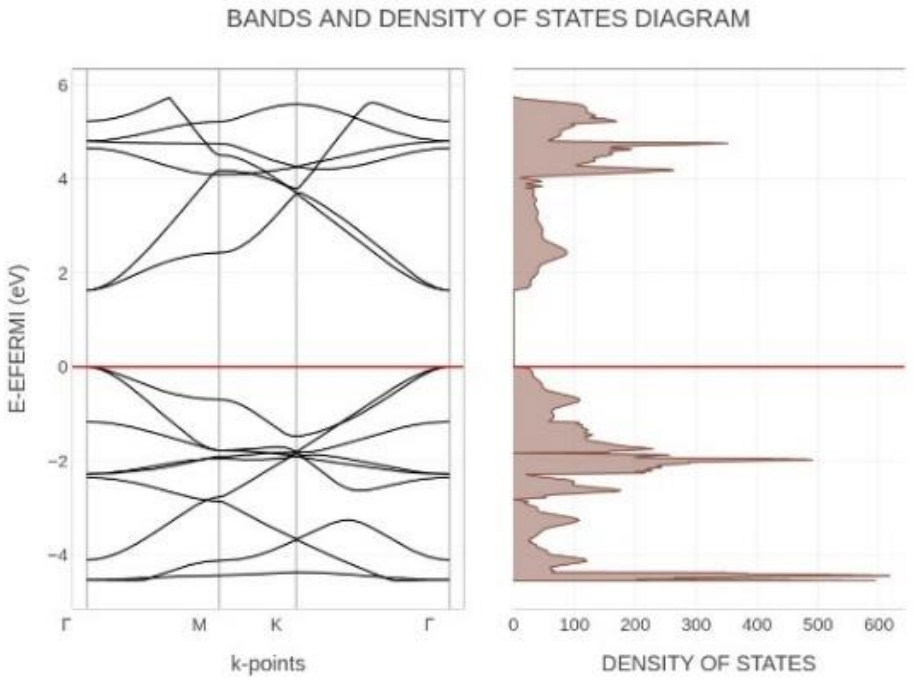</td>
<td>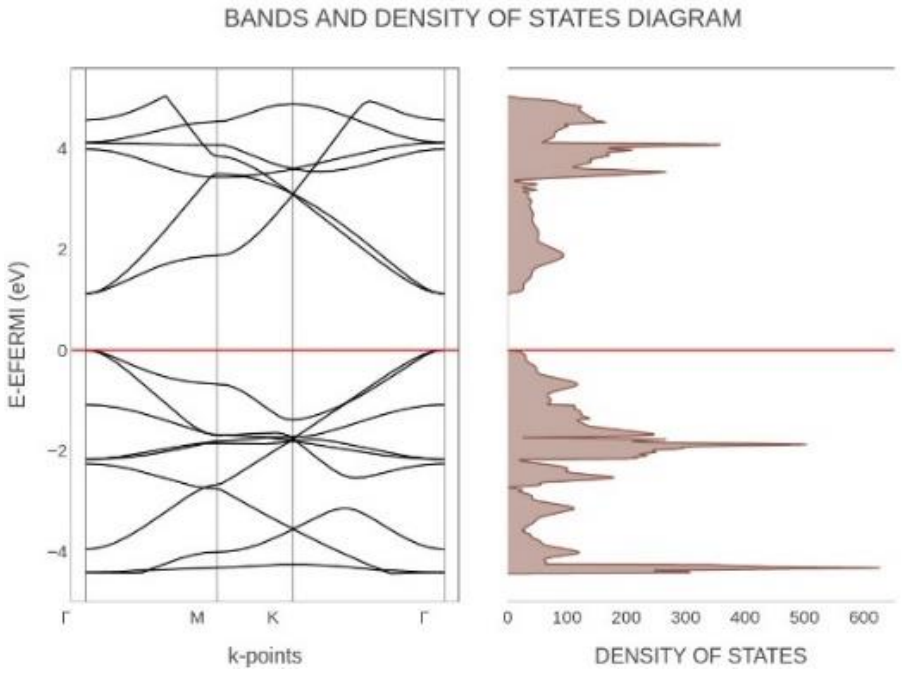</td>
</tr>
</table>

6-HT$^2$ (β-GDY)

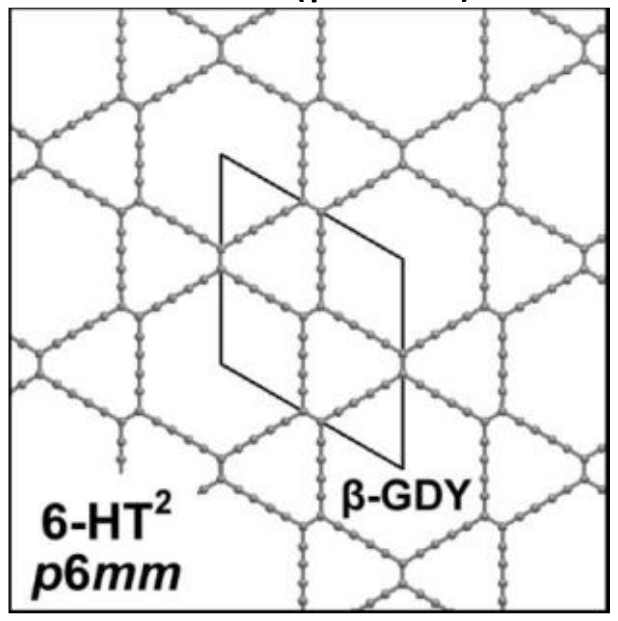

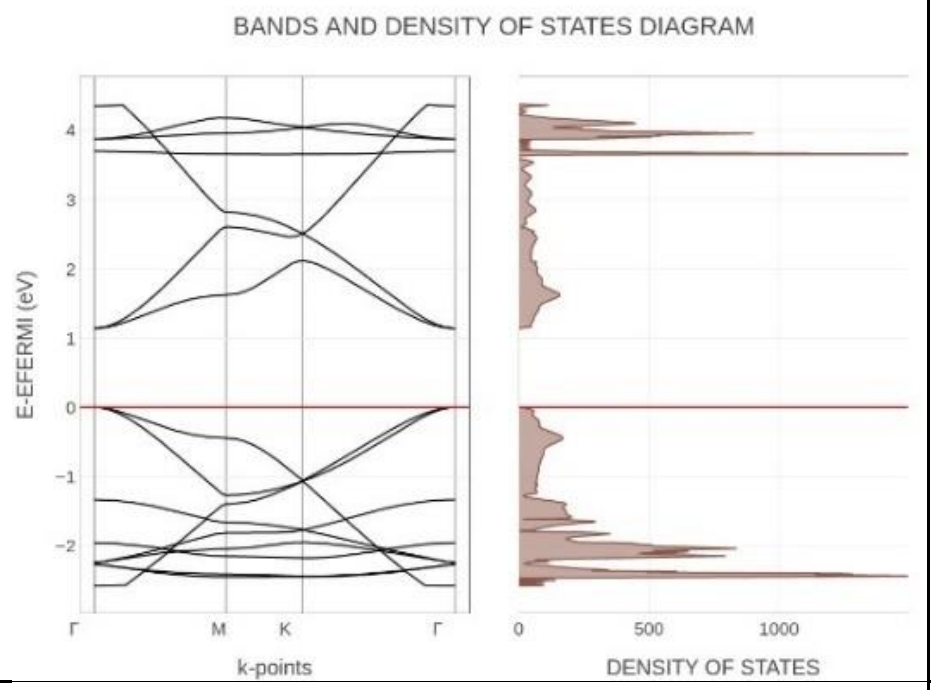

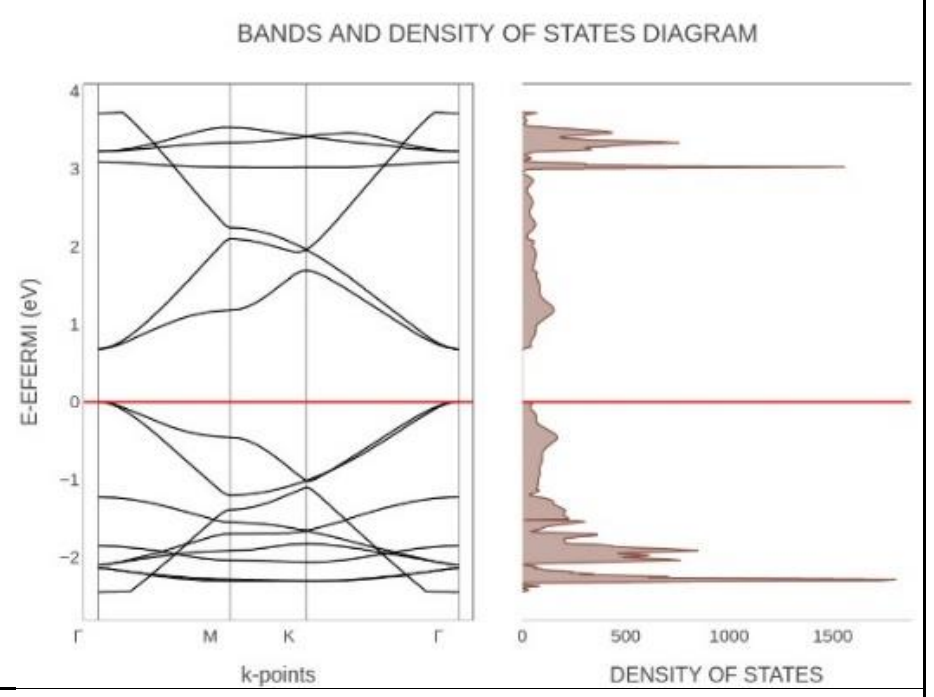

6-RL

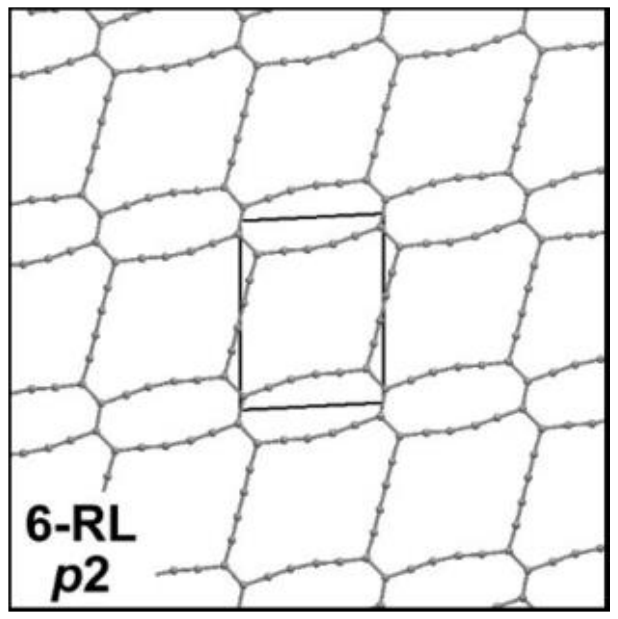

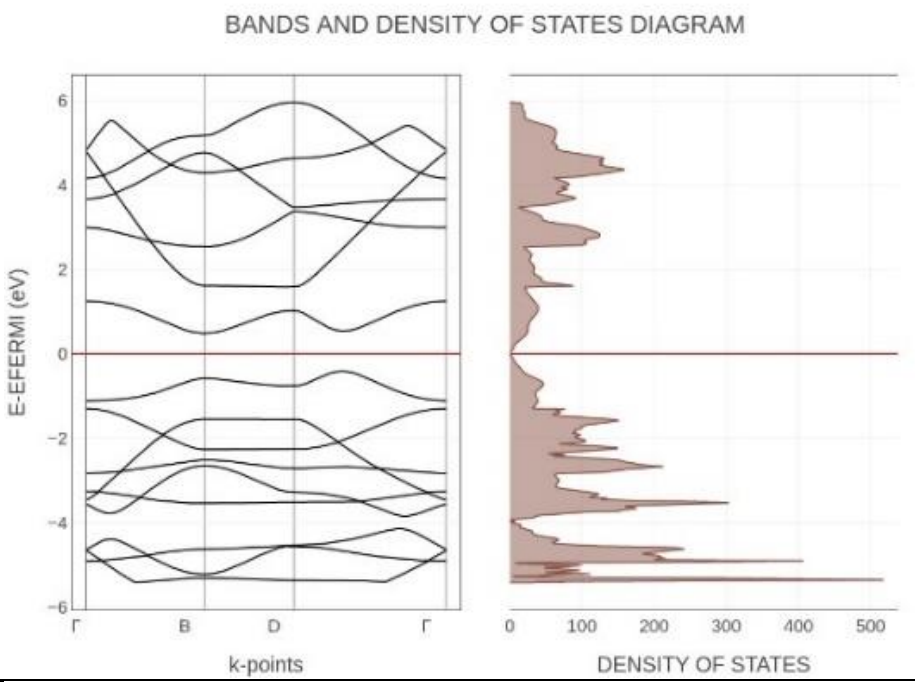

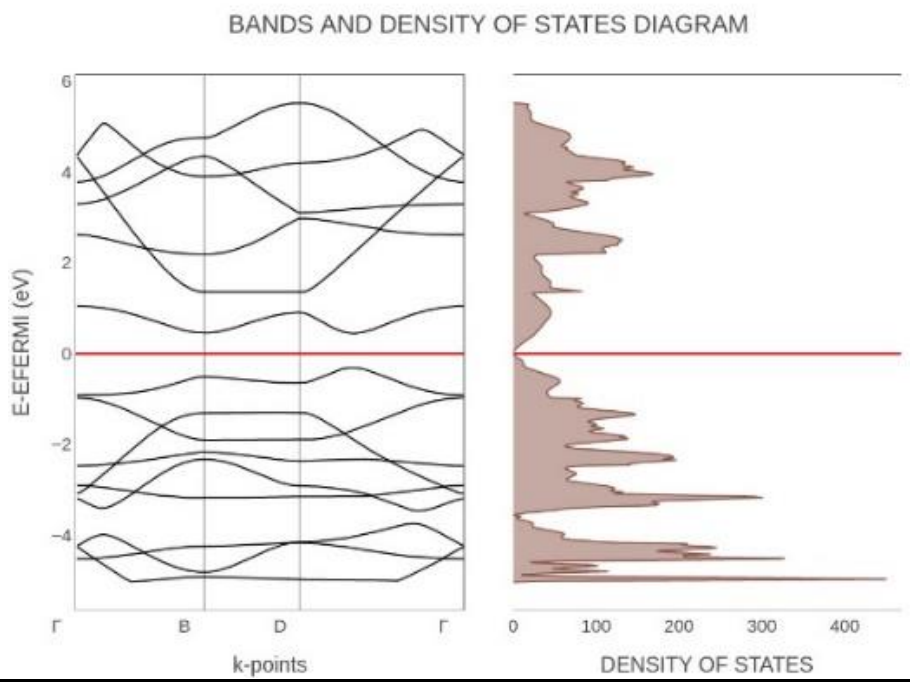

6-RL$^2$

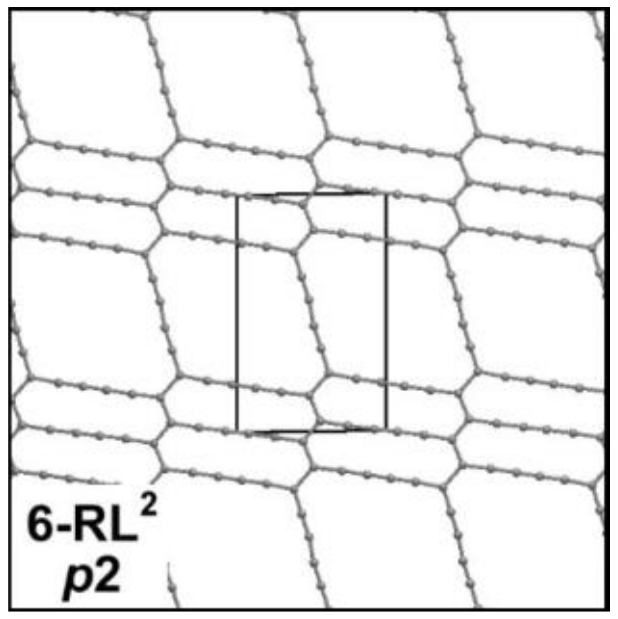

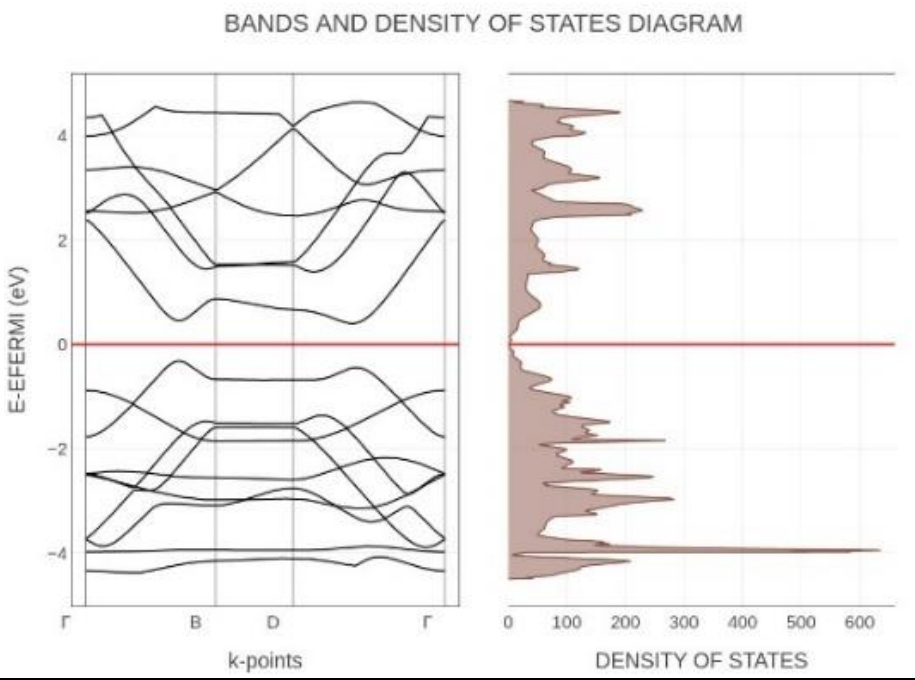

| | | |
|---|---|---|
| 6-$R^2$L<br>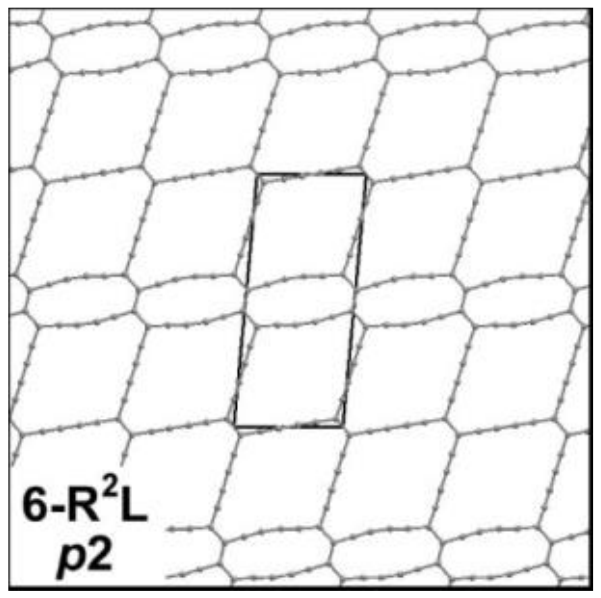 | 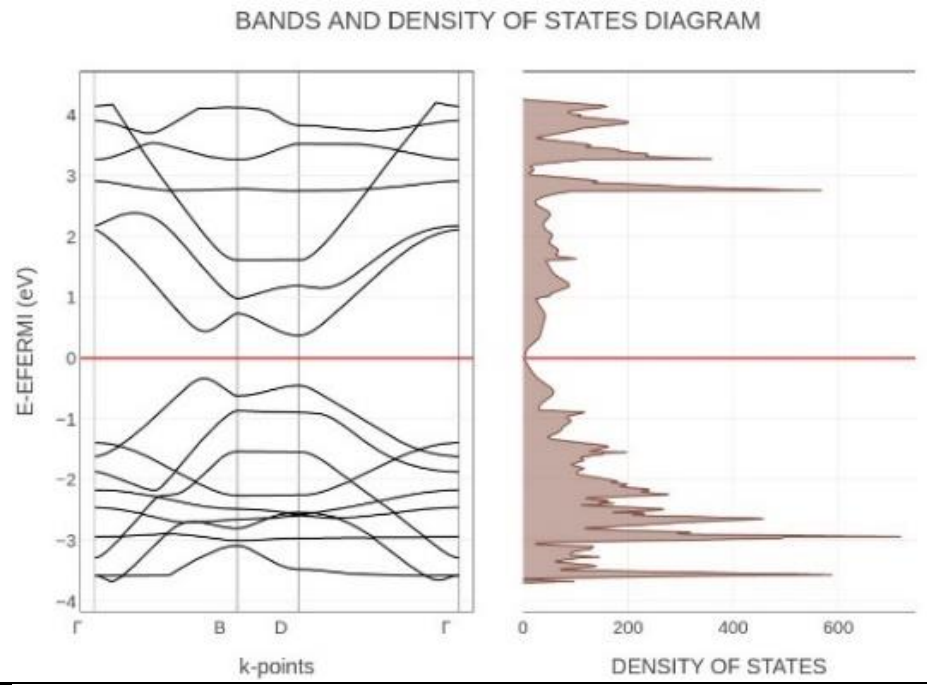 | |
| 6-H (α-GDY)<br>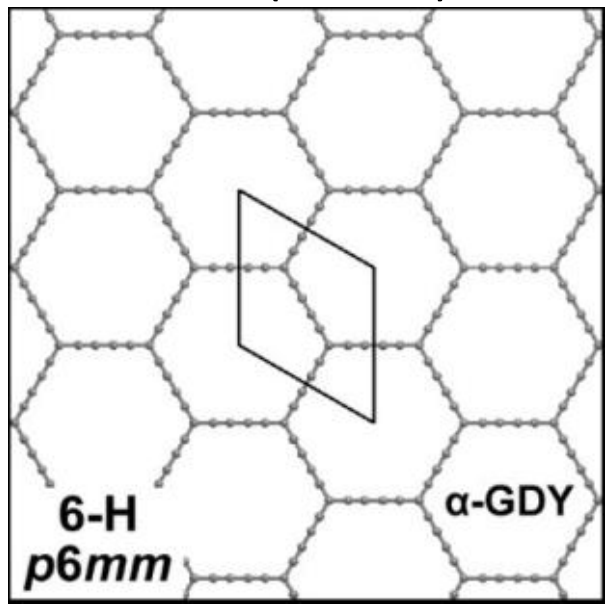 | 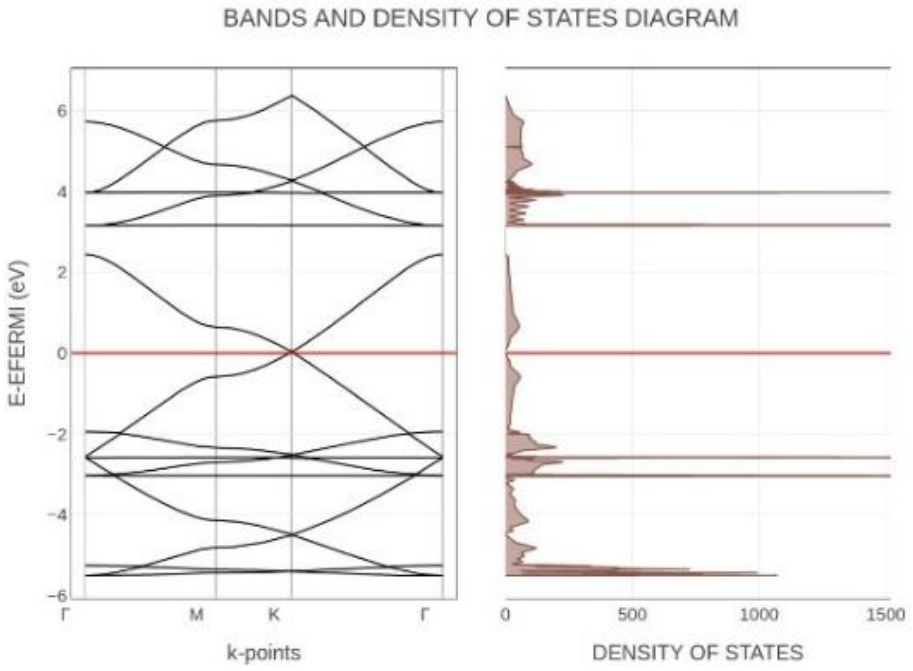 | 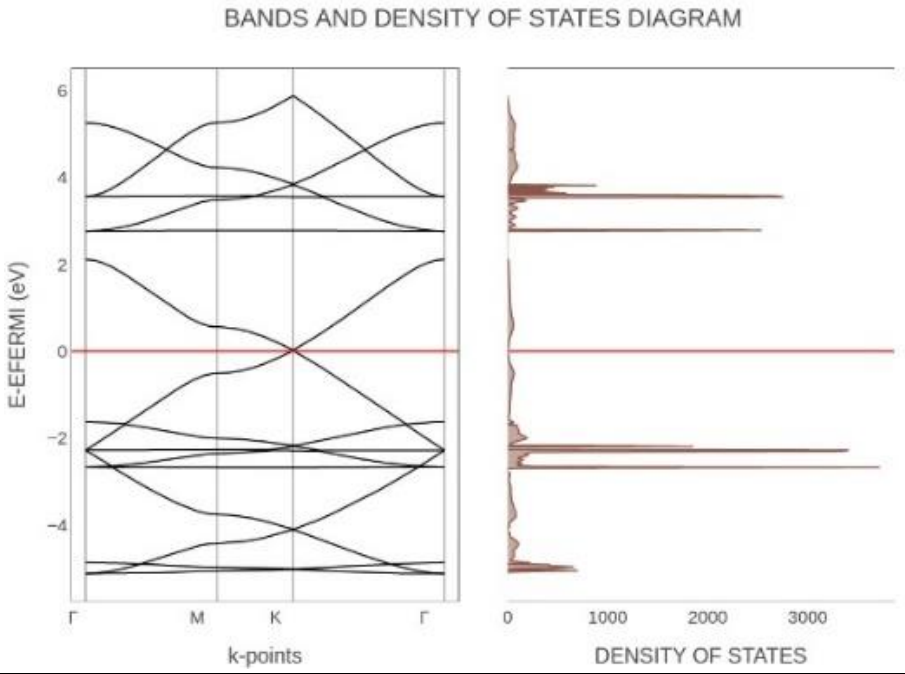 |

| # in Tab 1 | Names in red the new ones | Rel Energy kcal/mol | Area (Graphene 5.26Å$^2$) | density atomi/angs^2 graphene 0.380 | SPGR | Plane group | C(sp2)m X(sp)n | ratio sp/sp2 | Z | Pearson Symbol | # of C sp2 in the primitive cell | RINGS in the primitive cell | Rel Energy kcal/mol | Other names | JETP 2015 JPCC 2019 grazynes | npj Comp Mat 2018 | Phys. Status Solidi RRL 2020 | J Chem Phys 1987 |
|---|---|---|---|---|---|---|---|---|---|---|---|---|---|---|---|---|---|---|
| 1 | **6-h$^3$L** | **14,15** | 34,5 | 0,348 | Pmmm | p2mm | C X2 | **0,5** | 4 | oP12 | 8 | [6]$_3$[14] | **14,15** | | **[3],[2]-grazyne = 4-h$^3$L** | | | |
| 2 | **6-h$^2$L** | **16,81** | 58,2 | 0,344 | Cmmm | c2mm | C2 X3 | **0,66** | 4 | oS20 | 6 | [6]$_2$[14] | **16,81** | | **[2],[2]-grazyne = 4-h$^2$L** | | | |
| 3 | **6-h$^2$L$^2$** | **20,55** | 48,6 | 0,329 | Pmma | p2mg | C X | **1** | 8 | oP16 | 8 | [6]$_2$[14]$_2$ | **20,55** | | | | | |
| 4 | **6-hL** | **20,59** | 24,3 | 0,330 | Pmmm | p2mm | C X | **1** | 4 | oP8 | 4 | [6][14] | **20,59** | | **[1],[2]-grazyne = 4-hL** | | | |
| 5 | **6-hT$^2$** | **21,13** | 77,1 | 0,233 | P6/mmm | p6mm | C2 X | **2** | 6 | hP18 | 6 | [6][18]$_2$ | **21,13** | **γ-GDY graphdyne** | **γ1-graphyne-2** | **6-hT$^2$** | **191-E6Y12-1** | **6,6,6- = 4-hT$^2$ 6,6,6-GDY** |
| 6 | **6-hRL$^2$** | **22,65** | 191,9 | 0,250 | Cmmm | c2mm | C2 X | **2** | 16 | oS48 | 8 | [6][14]$_2$[22] | **22,65** | | | | | |
| 7 | **6-hT$^2$R** | **22,69** | 131,1 | 0,214 | Pmmm | p2mm | C5 X2 | **2,5** | 4 | oP28 | 8 | [6][18]$_2$[22] | **22,69** | | **β2-γ2-graphyne = 4-hT$^2$R** | **6-hT$^2$R** | **47-E8Y20-0** | **6,6,12- = 4-hT$^2$R 6,6,18-GDY** |
| 8 | **6-hL$^2$** | **23,11** | 87,3 | 0,321 | Cmmm | c2mm | C4 X3 | **1,33** | 4 | oS28 | 6 | [6][14]$_2$ | **23,11** | | | **>25Å** | | |
| 9 | **6-hL$^3$** | **24,06** | 63,1 | 0,317 | Pmmm | p2mm | C3 X2 | **1,5** | 4 | oP20 | 8 | [6][14]$_3$ | **24,06** | | | | | |
| 10 | **6-hR$^3$** | **24,07** | 166,5 | 0,192 | P6/mmm | p6mm | C3 X | **3** | 8 | hP32 | 8 | [6][22]$_3$ | **24,07** | | | **6-hR$^3$** | **191-E12Y12-1** | |
| 11 | **6-RL$^2$** | **24,73** | 87,2 | 0,252 | P2/m | p2 | C8 X3 | **2,66** | 2 | mP22 | 6 | [14]$_2$[22] | **24,73** | | | | | |
| 12 | **6-RL** | **24,73** | 72,7 | 0,220 | P2/m | p2 | C3 X | **3** | 4 | mP16 | 4 | [14][22] | **24,73** | | | | | |
| 13 | **6-R$^2$L** | **25,03** | 127,2 | 0,204 | P2/m | p2 | C10 X3 | **3,33** | 2 | mP26 | 6 | [14][22]$_2$ | **25,03** | | | | | |
| 14 | **6-HT$^2$L** | **25,07** | 202,4 | 0,178 | Pmmm | p2mm | C7 X2 | **3,5** | 4 | oP36 | 8 | [14][18]$_2$[30] | **25,07** | **δ-GDY** | **β1-γ2-graphyne = 4-HT$^2$L** | **[4-HT$^2$L]** | | |
| 15 | **6-HT$^2$** | **25,4** | 184,6 | 0,163 | P6/mmm | p6mm | C4 X | **4** | 6 | hP30 | 6 | [18]$_2$[30] | **25,40** | **β-GDY** | **β1-graphyne-2** | **6-HT$^2$** | **191-E6Y24-1** | **12,12,12- = 4-HT$^2$ 18,18,18-GDY** |
| 16 | **6-HL$^3$** | **25,41** | 165,6 | 0,193 | P6/mmm | p6mm | C3 X | **3** | 8 | hP32 | 8 | [14]$_3$[30] | **25,41** | | | **[4-HL$^3$]** | **191-E8Y24-1** | |
| 17 | **6-R** | **25,48** | 109,4 | 0,183 | Cmmm | c2mm | C4 X | **4** | 4 | oS20 | 2 | [22] | **25,48** | **carbon ene-yne CEY** | **β2-graphyne-2** | **6-R** | **65-E4Y16-0** | **14,14,14- = 4-R 22,22,22-GDY** |
| 18 | **6-HR$^2$L** | **25,85** | 474,9 | 0,168 | Cmmm | c2mm | C4 X | **4** | 16 | oS80 | 8 | [14][22]$_2$[30] | **25,85** | | | **>25Å** | | |
| 19 | **6-HR$^3$** | **26,11** | 276,1 | 0,159 | Pmmm | p2mm | C9 X2 | **4,5** | 4 | oP44 | 8 | [22]$_3$[30] | **26,11** | | | **>25Å** | | |
| 20 | **6-L** | **26,18** | 38,9 | 0,309 | Cmmm | c2mm | C2 X | **2** | 4 | oS12 | 2 | [14] | **26,18** | | **γ2-graphyne-2** | | **65-E4Y8-0 γ2-graphdiyne** | |
| 21 | **6-HR$^2$** | **26,3** | 442,5 | 0,154 | Cmmm | c2mm | C14 X3 | **4,66** | 4 | oS68 | 6 | [22]2[30] | **26,30** | | **α-β2-graphyne Fig 4b = 4-HR$^2$** | **>25Å** | | |
| 22 | **6-HR** | **26,63** | 166,7 | 0,144 | Pmmm | p2mm | C5 X | **5** | 4 | oP24 | 4 | [22][30] | **26,63** | | **α-β2-graphyne Fig 4a = 4-HR** | **6-HR** | **47-E4Y20-0** | **14,14,18- = 4-HR 22,22,30-GDY** |
| 23 | **6-H$^2$R$^2$** | **26,63** | 333,5 | 0,144 | Pmma | p2mg | C5 X | **5** | 8 | oP48 | 8 | [22]$_2$[30]$_2$ | **26,63** | | | **>25Å** | | |
| 24 | **6-H$^2$R** | **26,92** | 558,2 | 0,136 | Cmmm | c2mm | C16 X3 | **5,33** | 4 | oS76 | 6 | [22][30]$_2$ | **26,92** | | | **>25Å** | | |
| 25 | **6-H$^3$R** | **27,05** | 391,5 | 0,133 | Pmmm | p2mm | C11 X2 | **5,5** | 4 | oP52 | 8 | [22][30]$_3$ | **27,05** | | | **>25Å** | | |
| 26 | **6-H** | **27,39** | 112,5 | 0,124 | P6/mmm | p6mm | C6 X | **6** | 2 | hP14 | 2 | [30] | **27,39** | **α-GDY** | **α-graphyne-2** | **6-H** | **191-E2Y22-0** | **18,18,18- = 4-H 30,30,30-GDY** |

***Table S2****:* Crystallographic informations for all the 26 2D GDY-based crystals investigated in the work